\documentclass[%
reprint,
 amsmath,amssymb,
 aps,
prd,
]{revtex4-1}

\usepackage{graphicx}
\usepackage{dcolumn}
\usepackage{bm}
\usepackage{mathtools}
\usepackage{hyperref}
\usepackage{natbib}
\setcitestyle{square}

\begin{document}
\bibliographystyle{unsrt}
\newcommand{\mbf}[1]{\mathbf{#1}}


\preprint{APS/123-QED}

\title{ Causality Aspects of Modified Kerr-Newman spacetimes }

\author{Vaishak Prasad}
 \email{vaishak@iucaa.in}
\affiliation{%
 Inter-University Center for Astronomy and Astrophysics (IUCAA), Pune, India  411007\\
}%


\author{Rahul Srinivasan}
\email{f2013567@hyderabad.bits-pilani.ac.in }
\affiliation{
 Birla Institute of Technology and Science,Pilani - Hyderabad Campus\\
 Hyderabad, India 500078
}%
\author{Sashideep Gutti}
\email{sashideep@hyderabad.bits-pilani.ac.in}
\affiliation{%
 Birla Institute of Technology and Science,Pilani - Hyderabad Campus\\
 Hyderabad, India 
}%


\date{\today}

\begin{abstract}
In this paper, we address the problem of causality violation in the solutions of Einstein equations and seek possible causality restoration mechanisms in modifed theories of gravity. We choose for the above problem, the causality violation due to the existence of closed time-like curves in the context of Kerr-Newman black hole. We first revisit and quantify the details of the causality violation in the Kerr-Newman spacetime. We then show that the issue is also existent in two of the modified solutions to the Kerr Newman spacetime: The Non-Commutativity inspired solution and the f(R)-Gravity modifed solution. We explore the possibility of mechanisms present within the model that prevent causality violation. We show that, in both the models, the model parameters can be chosen such that the causality violating region is eliminated. We argue that in the context of non commutativity inspired solution, the non commutativity parameter can be chosen such that the causality violating region is eliminated and the inner horizon is no longer the Cauchy horizon. We then discuss the geodesic connectivity of the causality violating region in both the scenarios and quantify the geodesics that have points in the causality violating regions. We also discuss the causal aspects of Kerr Newman deSitter/antideSitter spacetimes.

\end{abstract}

\pacs{Valid PACS appear here}
\maketitle


\section{\label{sec:intro}Introduction}

Contrary to what was generally assumed erstwhile, we have learnt significantly in the last couple of years (Especially since the detection of Gravitational waves \citep{abbott2016}) that BH's are astrophysical objects of great importance, which are in direct contact with observations. Black-holes are also testing grounds for a variety of theoretical ideas since they present various scenarios where General Relativity and Quantum Theory become equally significant.The thermodynamic properties of the black-hole and the microstate justification for the entropy of the black hole, information loss, etc. are the usual milestones which are expeced from any theory of Quantum Gravity. Although General Relativity has passed major observational tests that we have been able to perform so far, the classical theory is known to have two main issues leading to breakdown of physics. The most well known of these are the issues connected with the existece of singulities where some of the scalar invariants diverge at one or more points of sapce-time. However, we now understand that these are generally expected to be existant due to the limitations of the Classical theory and are hoped to be resolved in a complete theory of Gravitation that incorporates both Quantum mechanics and Gravity. Presently, there are several existing approaches for this program, including String Theory, Loop Quantum Gravity etc. There are also some approaches that postulate that spacetime itself is Non-Commutative at small scales. The general expectations from such a theory is that they would smear away the point-like singularities of the Classical theory and replace them with regularized regions where the space-time curvatures remain finite. Since we do not yet have such a consistent theory, we do not yet have certainity in predictions of physics in certain regions of space-times that are in causal contact with singularities. For e.g. the interiors of the Kerr family of solutions to the Einstein's field equations. As a result of this, such regions are also inaccessible to investigations through numerical relativity as the field equations cannot be solved at the singularities.

The other issue, which is not directly concerned with the existence of singularities in classical theory that needs attention, is the breakdown of causality due to the admission of Closed Time-like Curves (CTCs) in some of the solutions of Einstein's equations (e.g. Kerr-Newman space-time, Gott spacetime, Godel Universe, etc) \citep{hawkingandellis},\citep{novikov1997}. On these curves, Chornology i.e. sense of ordering in time is lost, leading to the loss of determinism. Similar to the singularity problem, the regions of spacetime that are in causal contact with those containing CTCs are "viscious sets", also are inaccessible to investigations of Numerical Relativity. This is because, a Cauchy problem or an initial value problem is not well defined in the acausal region. The boundary of such a region where a Cauchy problem is ill defined is known as the Cauchy horion. Though there have been multiple attempts towards resolving the issues concerned with the existence of singularity in Classical and Quantum theories of gravitation \citep{bojowald2005} \citep{Mbonye2005}, \citep{hossenfelder2010}, there seem to not be many that address the issue of existence of CTCs.  In this paper we address the issue of violation of causality in the most astrophysically relevant of examples out of the set of those that are known to be causality violating: the charged rotating black-holes of the Kerr-Newman (KN) family. We  explore models of KN black holes in other modified solutions and seek possible mechanims within the model that can restore causality. The KN black hole provides a platform where both the singularity as well as the chronology preservation issue is present. The KN spacetime is therefore a good testing ground for theories of Quantum Gravity.

There are multiple issues concerning the existance of CTCs in the KN ST that make investigations of physics in the region difficult. The first being that the region where  CTCs occur (CTC region) lies in between the inner horizon of the KN black hole and the naked singularity\footnote{In some cases also extends 'beyond' the singularity to negative radii for e.g. Kerr ST, which may be considered physically irrelevant}. The CTC region in the KN space time is in causal contact with the naked singularity of the KN black hole, making the any data defined in the region corrupt. Thus the evolution of Cauchy data defined over the region with a naked singularity is an ill-defined problem \citep{hawkingandellis}. Owing to this, the pathology due to the CTCs was not required to be addressed and the region is not usually included in numerical investigetions and simulations. However, the second reason comes from a fundamental result in black hole perturbation theory. It was found in  that the inner horizon of the Reissner-Nordstrom and Kerr solutions was unstable (\citep{novikov1997} and the references therein). For the case of KN spacetime, as the inner horizon is also the Cauchy horizon, this implied the instability of the Cauchy horizon. It is argued that due to the infinite blue shift experienced at the Cauchy horizon, the sensibility of extending the manifold beyond the inner horizon should be questioned. This instability has been used to provide arguments in favor of the Strong Cosmic Censorship hypothesis which forbids local nakedness of singularity\citep{novikov1997}. Howeve in, \citep{cardoso18} it is  suggested that in a few situations the Cauchy horizon might not have the infinite blue shift as expected, and also that infinite blue shift is not enough to render the Cauchy horizon unstable and therefore the need to extend the manifold beyond the inner horizon.

 One main argument as to why the inner region of the KN blackhole and it's features are to be taken seriously is that if we consider a realistic scenario where the black hole is formed by the process of gravitational collapse. We do not yet have an analytical model of collapsing matter whose final state is a Kerr-Newman black hole. If we take the hints from the gravitational collapse that leads to the formation of Schwarzschild black hole, one can have various scenarios based on the initial configuration of the collapsing matter \cite{pankaj2011}, \cite{malafarina2017}. In some situations, the apparent horizon forms before a singularity and yet in other collapsing scenarios one can have the singularity forming before the formation of apparent horizon etc. So if a similar scenario is true for the formation of K-N blac khole, then features like the CTC region might emerge before the formation of singularity or the apparent horizons, etc. The progress towards understanding the dynamical formation of K-N blackhole is made in the numerical study \cite{nathanail17}.  Though no conclusion can be reached regarding the cauality violating region. At this stage of current research, our guiding principles wrt the collapsing scenarios are Cosmic Censorship Hypothesis (both weak and strong) and Chronology Protection Conjecture.

We will begin by stating the Chronology Protection Problem and illustrate it in the context of the Kerr-Newman black-hole of Einstein's Gravity. Here we describe the existence of CTC in the manifold and explain how it leads to the breakdown of causality. We then address the Chronology Protection problem in two popular modified solutions for Kerr-Newman spacetime: The Non-Commutativity inspired Kerr-Newman solution, one of the modified models of BH that has effectively avoided the point-like singularities, and the Kerr-Newman solution of $f(R)$ gravity. In both cases, we first investigate the existence and basic properties of the CTC region and analyze null geodesics in order to understand how well the region containing closed time-like curves is causally connected with the outer regions. We then seek to find possible natural mechanisms within the models that can preserve causality.


\subsection{\label{sec:CPC}The Chronology Protection Conjecture and its violation}

The Chronology Protectection Conjecture was first given by Hawking in 1992 amidst the growing number of suggestions that one maybe able to time-travel if he/she travels faster than speed of light \citep{hawking91}. He stated that:

\emph{"The 1aws of physics prevent the appearance of closed time-like curves".}

Closed time-like curves (CTCs) are said to exist in a space-time if there exists a directed time-like vector field defined in a region whose integral curves are closed. It is obvious that such curves pose a threat to the principle of causality. The world like of a particle in circular motion about a black hole, for instance,  does not qualify to be a CTC since the circular ourbit is just its spatial trajectory. The temporal translation ensures that the world line is helical. 

Solutions of Einstein's equations containing CTCs were first found by Godel\cite{godel1949}. Numerous other solutions were found later. Time-travel was being considered as a theoretical possibility and questions were being raised in order to find out if space-time can be curved so much as to bend the light cone structure of space-time sufficiently in a region creating CTCs. In the region of space-time where CTCs exist, the local light-cone structure is tipped along the closed worldline of the observer all along the curve. Thus, although locally the space-time manifold is still Lorrentzian, the existence of CTCs is a gloabal property of spacetime. These regions containing CTCs are therefore expexted to exist near very high curvatures of spacetimes.  Due to this requirement, the existence of CTCs require a constraint on the type of matter fields that can support and maintain the CTCs in the region.  It was shown that either the \emph{Weak Energy Condition} is violated or singularities exist in such regions \citep{hawking91}\citep{morris1988}

\begin{align}
T^{ab}l_a l_b < 0, 
\end{align} 
where $T^{ab}$ is the Energy momentum tensor and $l_a$ is a future directed null vector field. 

It was shown by Carter in \citep{carter1968} that in the KN spacetime, the violation of causality due to the CTCs is inevitable and cannot be resolved by the choice of a suitable covering space.

Since a Cauchy problem is not well stated in the CTC region, we cannot ensure evolution consistent with the structure of the background spacetime as time is periodic. By considering the matter fields/ dynamical quantities of interest as test fields, restricting the initial conditions i.e. by imposing periodic boundary conditions on them , one can atmost ensure that the eternally periodic solutions are consistent with the background spacetime structure.
However, when these fields are coupled to the Einstein's equations, back-reaction effects are included and then it is not clear if such solutions ( eternally periodic) continue to be consistent. 

We presently know of only a few exact solutions to Einstein's Gravity, each of which presents us with valuable insights and brings to light an important feature of the theory itself. Among these are also solutions in which causality can apparently be violated. The well known examples to these are the Gott spacetime, Godel universe, Kerr spacetimess, etc. The generic manner in which this appears is through the admittence of closed time-like curves by the spacetime. These problems are known to persist through maximal extensions of the manifold and cannot be gotten rid of by the choice of a suitable covering space \citep{carter1968}.  


We first revisit and summarize the details of causality violation in the Kerr Newman spacetime. The Kerr-Newman metric has the following form in Boyer-Lindquist coordinates:
\begin{align}
ds^2 = -& \dfrac{(\Delta - a^2sin^2\theta)}{\rho^2} dt^2 - \dfrac{2asin^2\theta}{\rho^2}(2M r - q^2)dtd\phi \\+& \dfrac{\rho^2}{\Delta}dr^2 + \rho^2 d\theta^2 + \dfrac{\Sigma^2sin^2\theta}{\rho^2}d\phi^2 \label{KNmetric}
\end{align}

Where,
\begin{align}
\Delta = \:& r^2 - 2M r + a^2 + q^2 \\
\Sigma^2 = \:& (r^2+a^2)^2-a^2\Delta sin^2\theta\\
\rho^2 = \:& r^2 + a^2cos^2\theta
\end{align}

Here $M$ is the mass of the Black hole, $a$ its specific angular momentum and $q$ is the charge of the blackhole.
The classical features of Kerr Newman metric include the existence of two horizons, ergosphere and the ring singularity.  The two horizons which are the roots of the equation $\Delta=0$ are as given below,

This has the interpretation of representing the spacetime of a charged,rotating Black-hole. Although one does not know whether this also represents the exterior metric of a finite sized charged,rotating object, it is known that all such metrics must be asymptotically Kerr. Moreover, the uniqueness theorem guarantees that the metric of a charged rotating BH is Kerr-Newman. Thus, it is of significance to study the KN spacetime.We now briefly recap the important features of the KN spacetime.

The classical features of Kerr-Newman metric include the existence of two horizons, ergosphere and the ring singularity.  The two horizons which are the roots of the equation $\Delta=0$ are as given below,

\begin{align}
r_{-}=& \mu - \sqrt{\mu ^2 - (a^2+q^2)}\\
r_+= &\mu + \sqrt{\mu ^2 - (a^2+q^2)}
\end{align}

which are the inner and outer horizons respectively.

 These exist only when 

\begin{align}
\mu^2 \geq a^2+q^2
\end{align}

The quantity $ \Delta - a^2sin^2\theta$ is negative at the horizons where $\Delta = 0$.
The location of the singularity is again at $ \rho =0$ :

\begin{align}
r=0 \qquad cos\theta = 0
\end{align}
which corresponds to a ring of radius $a$ in the equatorial plane. 

It is quite well known that the existence of the horizon and the ergo-region lead to interesting physics ( like Hawking radiation, Superradiance, QNMs etc.) 
\begin{figure*}[htbp]
\includegraphics[width=0.49\textwidth]{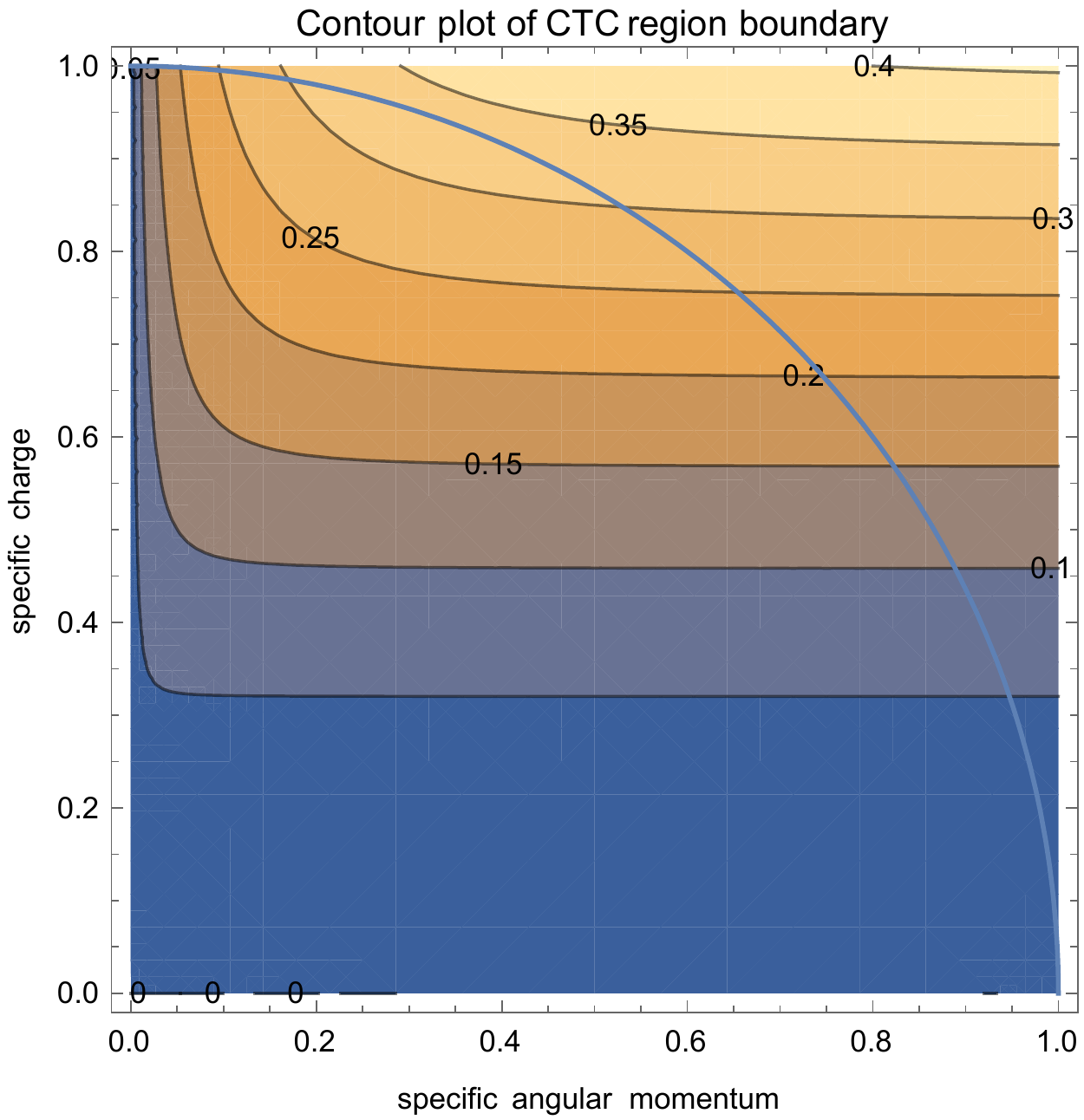}
\includegraphics[width=0.49\textwidth]{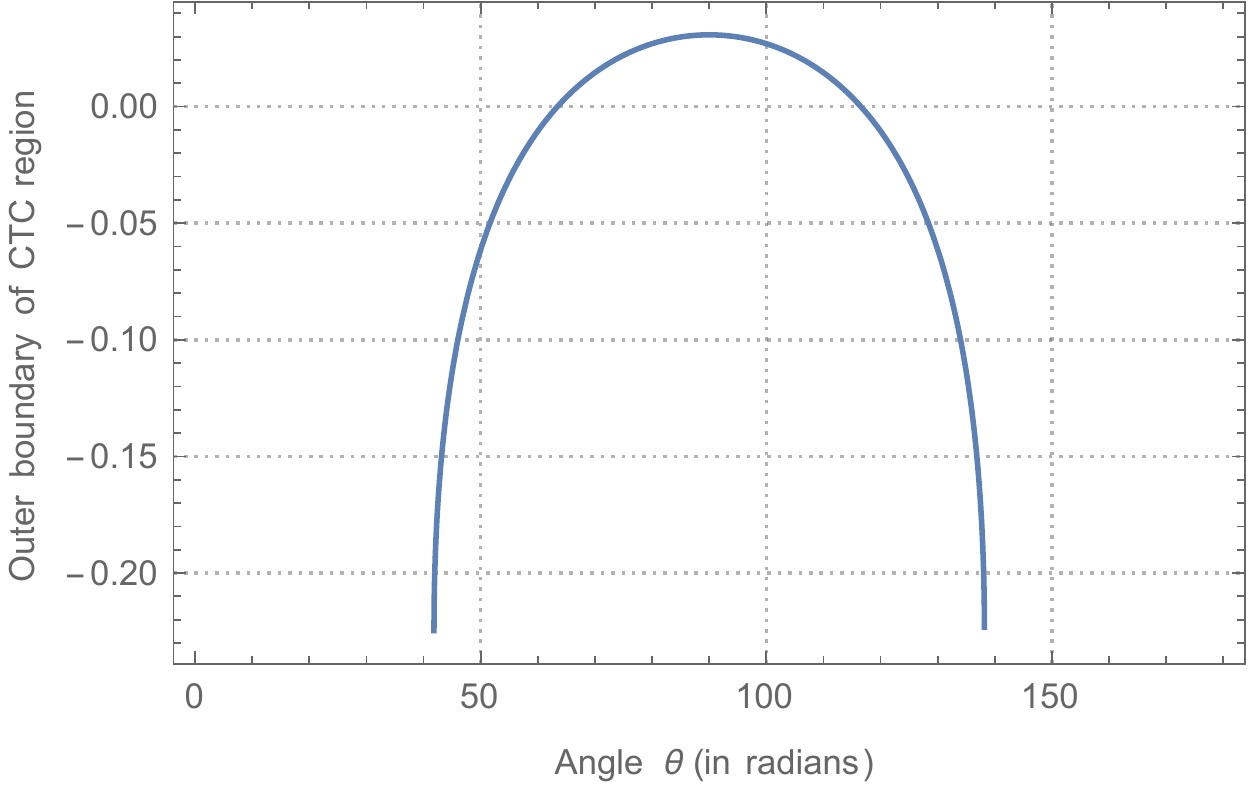}
\caption{The left panel shows a contour plot of the outer radius of the CTC region on the equatorial plane ($\theta = \pi/2$) against the BH parameters, normalised by the BH mass ( $a' = a/M$ and $q' = q/M$ here). These contours are labelled by the radius of the region containing closed time-like curves. Also superimposed is the boundary of the region $a'^2 + q'^2 = 1$, a semicircle. All the points lying inside the region are physically probable end states of a collapse corresponding to the condition $M^2 > a^2 + q^2$. Note that there are BHs corresponding to these end states that harbour a CTC region. Right: Plot of outer radius of CTC region vs. the Kerr-Schild anglular co-ordinate $\theta$ ( measured from the spin axis).}
\label{fig:KNcont}
\end{figure*}

 \emph{Closed time-like curves} are closed curves whose tangent vectors are everywhere future directed time-like/ past directed time-like. Such curves pose a threat to the validity of causality as the chronological future of a point $\mbf{x}$ on the curve may also contain points from its chronological past:
\begin{align}
I^+(\mbf{x}) \cup I^-(\mbf{x}) \neq \emptyset \label{chint}
\end{align}

Consider the axisymmetric vector field $\xi_{(\phi)}^\mu$ (which is also a Killing vector). The norm of the vector field is given by $\xi_{(\phi)}^2 = g_{\phi\phi}$. Now for simplicity, consider the metric component $g_{\phi\phi}$ (algebraically simplified) on the equatorial plane ($\theta = \pi/2$):

\begin{align}
g_{\phi\phi} = r^2 + a^2 + \dfrac{2Ma^2}{r} - \dfrac{a^2 q^2}{r^2}
\label{KNgppsimp} 
\end{align}
Here we can see that the first three terms are strictly positive whereas the last one is strictly negative. Also, we see that the last term dominates at smaller radii.

This shows that $g_{\phi\phi}$ may not have the same sign everywhere on the manifold. It can change sign at the radius $r_{ctc}$ where $g_{\phi\phi}$ vanishes. By analyzing the nature of the roots of the equation through their sum and products, we can arrive at the fact that there are two real roots, out of which one of them is positive and we call this $r_{ctc_+}$. We will address the other smaller root by $r_{ctc_-}$. Thus in the region where $g_{\phi\phi}$ takes negative values, we find that the norm $\xi_{(\phi)}^2$ is negative:

\begin{align}
\mbf{\xi_{(\phi)}}\cdot \mbf{\xi_{(\phi)}} = g_{\phi\phi} < 0  
\end{align}

This shows that the integral curves of $\frac{\partial}{\partial \phi}$ can become \emph{time-like} in nature. It is hence clear that the co-ordinate $\phi$ becomes time-like in the region $r_{ctc_-} < r < r_{ctc_+}$. Furthermore, since the integral curves of $\frac{\partial}{\partial \phi}$ are closed (topologically $S^1$), there exist time-like curves that are \emph{closed} \emph{in the full 4-dimensional space-time} \citep{mthesis}. The periodicity in $\phi$ has led to the periodicity of time for observers whose world line is one of the closed time-like curves. This means that one can have a sequence of events in the ST that close back onto themselves. Thus the observer may violate causality.
It is evident from the above equation \eqref{KNgppsimp} that closed time-like curves also exist in the Kerr spacetime, however at small negative radii. The presence of the Electromagnetic field and the manner in which it couples to Einstein's gravity has somehow been reponsible in bringing this region to positive radii.

Due to the equivalence principle, one can establish a locally flat coordinate system in which the metric is approximately Minkowski, in which we know that there is no causality violation. The light cone-structure of the space-time is locally preserved. Therefore we understand that the possibility of violation of causality is not a local property of the space-time, but rather a global property depending upon the manner in which space-time points are connected. Since the topology of integral curves of $\xi_{(\phi)}$ is a co-ordinate independent property we see that the existenc e of closed time-like curves is a global property of the space-time itself that is unavoidable \citep{carter1968}.

One can discern the placement and the shape of the region as follows. The condition $g_{\phi\phi} = 0$ on the equatorial plane $\theta = \Pi/2$ gives the following equation:

\begin{align}
h(r) = r^4 + a^2r^2 + 2M a^2 r - a^2q^2 = 0
\end{align}

As mentioned earlier, as long as $q \neq 0$, it can be reasoned out that this equation admits one positive and one negative root. Further, it can also be shown that the only positive root lies inside the inner horizon $r_c < r_{h_-}$ (See \citep{mthesis}). In the contour plot Fig. \ref{fig:KNcont} we show the dependence of $r_{ctc_+}$ on the BH parameters. It can be seen that for these chosen parameters, the existence of the CTC region is evident. The adjoinig figure shows a plot of the outer boundary of the CTC region 
(as defined by the outermost positive real root of $g_{\phi\phi} = 0$) vs.  $\theta$. This shows that the CTC region extends out farthermost on the equator, diminishing in size on either of its sides. Also, in the .

Given that CTCs exist in the KN ST, it is of importance to study how exposed the region containing these curves is. 
Significant amount of research has been done to analyze the CTC region, (\citep{carter1968} \citep{defelice1980} \citep{Slobodov2008} \citep{gonzalez1996} \citep{salazar17} \citep{boulware1992} \citep{mthesis}) showing that there exist null geodesics that can sample points from the CTC region. Carter presented a maximal extension of the KN BH and were first to show that existence of CTC and that it cannot be cured by means of a suitable covering space . They also showed that only those particle trajectories which are restricted to the equatorial plane can reach the ring singularity \citep{carter1968}. In \citep{mthesis} we presented a different analysis and re-derived some of these results. In particular, we showed that there exist a range of geodesic parameter values ${E,L,m_0}$ for which there exist turning points in the CTC region. Thus there are geodesics that sample the CTC region, carrying causality violating data to the outside observer. In this sense, the violation of causality in the KN ST is completely exposed.

In this paper, we analyze the validity of causality in modified KN metrics, namely the Non-commutativity inspired Kerr-Newman BH \citep{lmodesto10} and the Kerr-Newman Blackhole in f(R) gravity \citep{cembranos14}. 

\begin{figure*}[htbp]
\includegraphics[width=0.49\textwidth]{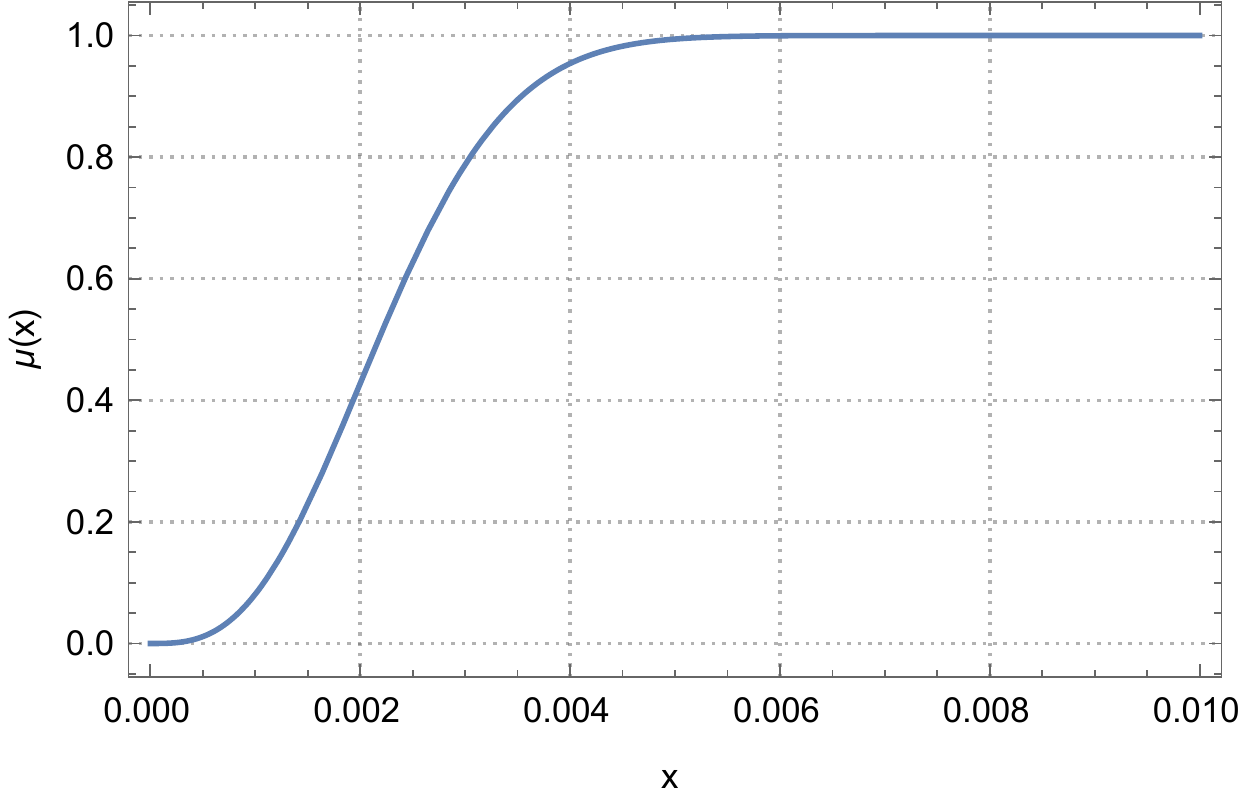}
\includegraphics[width=0.49\textwidth]{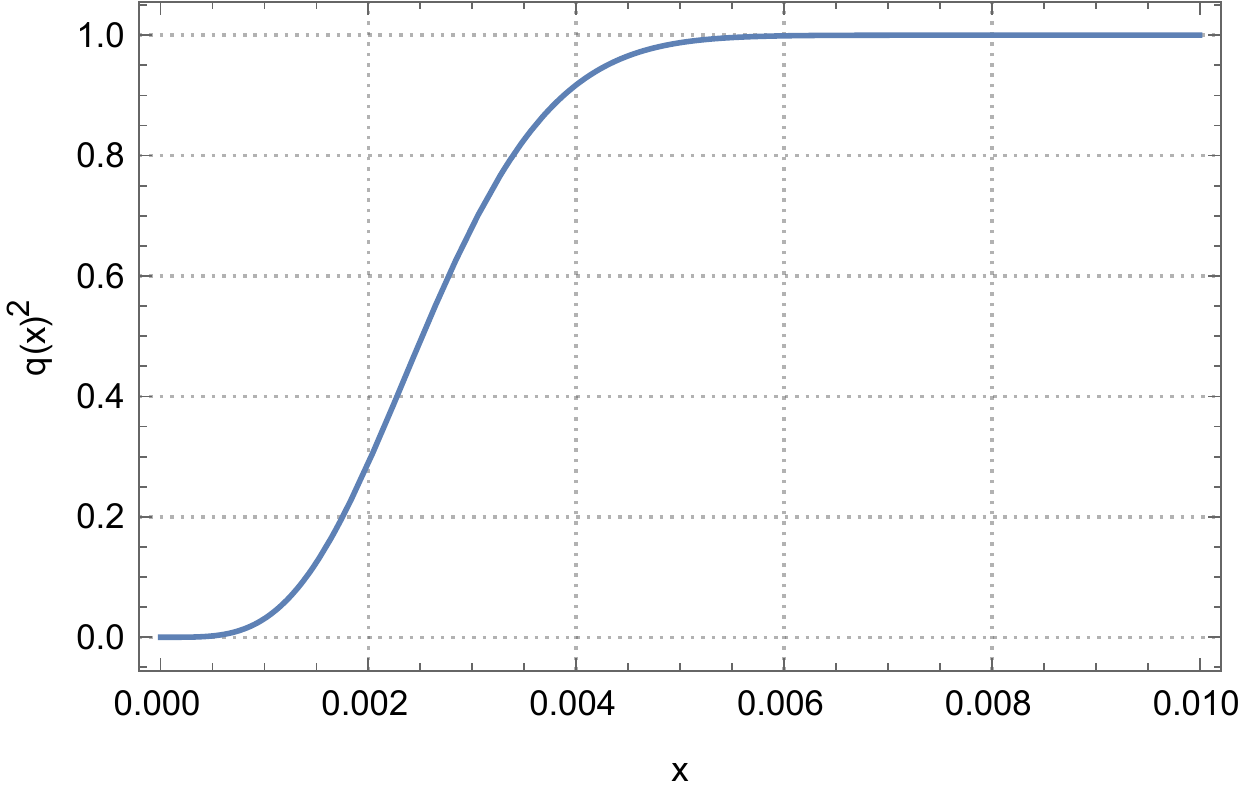}
\caption{The plot of unit mass and squared unit charge of a Non-Commutative Kerr-Newman BH with the normalized radius $x = r/M$ . The BH parameters in the Non-Commutativity inspired BH are no longer Dirac-delta distributed. They have a spread due to the Gaussian nature of their singularity. The Non-Commutative BH starts resembling its usual counterpart ( Kerr-Newman BH) at scales comparable/greater than the length-scale characterizing the spread of the singularity due to non-commutativity $\sqrt{p}$. Thus the BH parameters asymptotically attain their actual values of unity at $x>\sqrt{p}$.}
\label{fig:NCKNmuq}
\end{figure*}

\subsection{\label{sec:MC}Methods and Conventions}
We use semi-analytic methods employing Mathematica for numerical analysis and graphing. We use $G = c = 1$ units throughout. All the parameters and variables are expressed in mass units i.e. the BH mass $M$ is also set to $1$. Therefore, the normalized specific angular momentum and specific charge are related to their non-normalized quantities by $a' = a/M, q' = q/M$

\section{\label{sec:NCKNintro}Causality violation in Non-commutativity inspired Kerr-Newman spacetime}

Non-commutative (NC) Black-holes are Black-holes inspired by Non-Commutative Geometry. They incorporate a minimal length scale in the metric effectively induced by the non-commutative character of the co-ordinate operators \citep{lmodesto10}\citep{nicolini2006}\citep{ansoldi2007}\citep{spallucci2009}. These aim to cure the BH point-like curvature singularities (e.g. the Dirac-delta like singularity of Schwarzschild BH, the ring singularity of the Kerr BH) by effectively replacing it by a distribution. From the consideration of the NC nature of co-ordinate operators in the co-ordinates coherent state approach and expected first order quantum gravitational correction to classical General Relativity, one can show that the Dirac-delta like singularities will be replaced by Gaussian distributions (see \citep{lmodesto10}).

For instance, for a point particle at the origin, the non-commutativity implies that the Dirac-delta $\delta_D^3(\mathbf{x})$ is effectively replaced by:

\begin{align}
\rho_0(\mathbf{x}) = \dfrac{1}{\sqrt{2\pi p}} e^{-\mathbf{x}^2/2p} 
\end{align}

The width of the Gaussian distribution is characterized by the non-commutativity parameter $p$ having the dimensions of length squared, that encodes information about the minimal length scale in the non-commutative space-time. 

Furthermore, it has also been shown that the effective corrections to the Einstein's-field equations due to this can be modeled by only modifying the source term (the matter sector, i.e. the Energy-Momentum tensor), replacing the point like sources by a suitable Gaussian distribution while the differential operators (i.e. the geometry sector ) remain unchanged \citep{nicolini2005}. 

Following this approach, first the NC solution for the Schwarzschild BH was derived by Piero and Euro in 2005 \citep{nicolini2006}, followed by the NC Reissner-Nordstrom solution \citep{ansoldi2007}. Later, the NC solution for the rotating BH (Kerr) and the charged-rotating BH (Kerr-Newman) \citep{lmodesto10} were also obtained by means of a modified Newman-Janais algorithm. The Kerr family of NC BH incorporate generic Quantum Gravity effects and have also been extended to higher dimensions \citep{lmodesto10},\citep{spallucci2009}

The most important features of these NC models is the resolution of the singularity. The interior-most region is regular and contrary to the classical scenario, this region is not plagued by the corrupting data from the singularity. However, as mentioned earlier, the avoidance of singularity is not the only problem a complete theory of Gravity (that incorporates Quantum effects) should address. Classical Einstein's Gravity is also troubled with other major issues like the existence of closed time-like curves which have to be eventually addressed.

The NC KN metric has the form:
\begin{align}
ds^2 = -& \dfrac{(\Delta - a^2sin^2\theta)}{\rho^2} dt^2 \\ -& 2asin^2\theta \left(1-\dfrac{\Delta-a^2 sin^2\theta}{\rho^2}\right)dtd\phi\\ +& \dfrac{\rho^2}{\Delta}dr^2 + \rho^2 d\theta^2\\ +& sin^2\theta\left(\rho^2 +a^2sin^2(\theta)\left(2-\dfrac{\Delta-a^2 sin^2\theta}{\rho^2}\right)\right)d\phi^2 \label{NCKNmetric}
\end{align}

with 

\begin{align}
\Delta = \:& r^2 - 2M(r) r + a^2 + q(r)^2 \\
\rho^2 = \:& r^2 + a^2cos^2\theta
\end{align}
Here $M(r)$ and $q(r)^2$ are given by the following expressions.
\begin{align}
M(r)=\dfrac{\gamma(3/2;r^2/(4p))}{\Gamma(3/2)}
\label{mr}
\end{align}
\begin{align}
q(r)^2 = \dfrac{Q^2}{\pi}\Big[&\gamma^2(1/2;r^2/(4p))-\dfrac{r}{\sqrt{2p}}\gamma(1/2;r^2/(2p))\\ &+ r\sqrt{\frac{2}{p}}\gamma(3/2;r^2/(4p))\Big]
\label{NCKNsymb}
\end{align}

The source corresponding to the NCKN spacetime is non-zero (can be found in \citep{lmodesto10}). In particular, the $T^0_0$ component of the energy-momentum tensor $T_{\mu\nu}$, which is the source of the gravitational field given by \eqref{NCKNmertic}, is going to assume the form of a Gaussian as described by the equation above \eqref{NCKNsymb}.
The metric can be seen to be symbolically similar to its classical counterpart ( $g_{\phi\phi}$ also can be simplified and brought to the same form as that of KN \eqref{KNmetric}), but the parameters are now accompanied by a spread due to the NC nature of the space-time. Here the parameters $M$, $Q$ and $a$ become the ADM mass, charge and angular momentum of the black-hole at asymptotic region.  The Non-Commutative KN BH only deviates from its classical counter-part at small length scales ($l\sim \sqrt{p}$). At large length scales ($l>>\sqrt{p}$), the mass $M(r)$ and specific charge $q(r)$ approach the limiting values ($M,q$) respectively and the space-time is effectively the classical KN space-time.
We recap some of the main results of the paper \citep{lmodesto10} that are relevant for our present study.

\begin{itemize}
 \item The horizons are the roots of the equation $\Delta=0$. The quadratic equation is now replaced by a transcendental equation. The solution that yields a black-hole is when the equation has at least one positive root. The equation can have at most two positive roots. The root diagram for the Kerr Black-hole case  as a function of parameters can be found in \citep{lmodesto10}. Here we extend the analysis for the case of Kerr-Newman. This analysis helps us point out the physically relevant parameters that yield a black-hole.
 \item The metric and the curvature scalars are regular near origin. Also, the NC KN black-hole has a repulsive core. Further details can be found in \citep{lmodesto10}. 
 \item $g_{\phi\phi} \rightarrow a^2$ as $r \rightarrow 0$.  
 \end{itemize}

\subsection{\label{sec:NCKNCTC}The Closed Time-like curves region}

\begin{figure*}[htbp]
\includegraphics[width=0.49\textwidth]{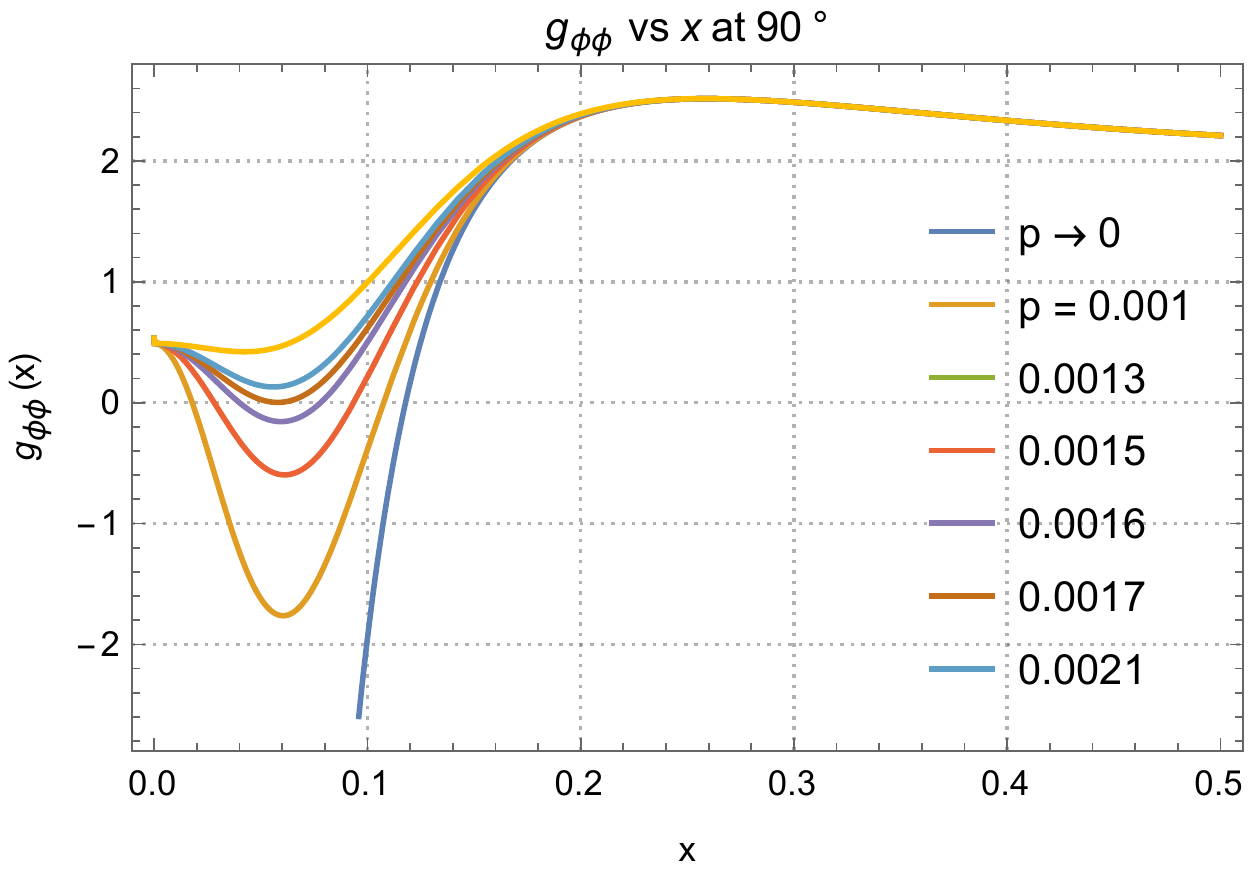}
\includegraphics[width=0.49\textwidth]{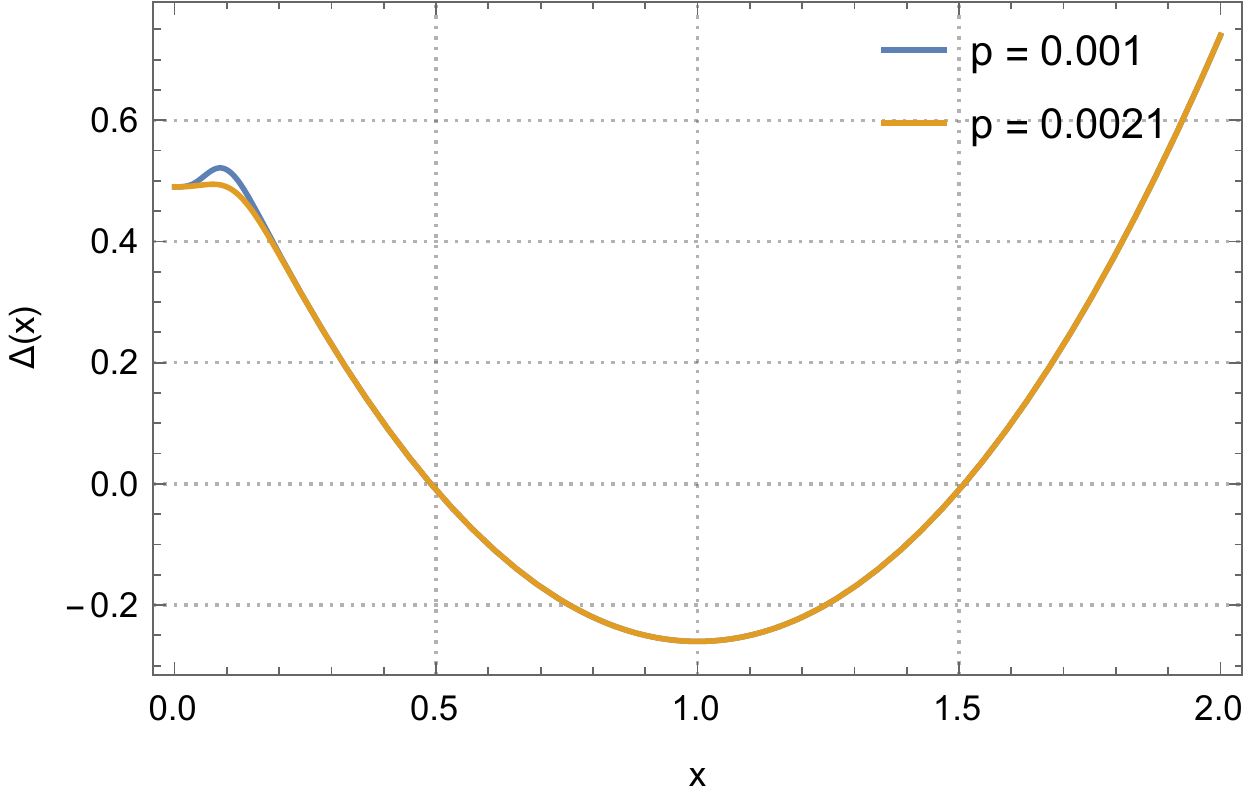}
\caption{The first panel shows the metric component $g_{\phi\phi}$ is plotted vs. $x =r/M$. Note that for sufficiently large value of $p$, the equation $g_{\phi\phi} =0$ has no positive real roots, and thus no CTC region.  The second panel shows the corresponding plot of the horizon function $\Delta(x)$ vs. x. The variation in $p$ only affects the graph at small scales comparable to $\sqrt{p}$. All the BHs corresponding to the above parameters have two horizons.}
\label{fig:NCKNgppxp}
\end{figure*}

\begin{figure}[htbp]
\includegraphics[width=0.49\textwidth]{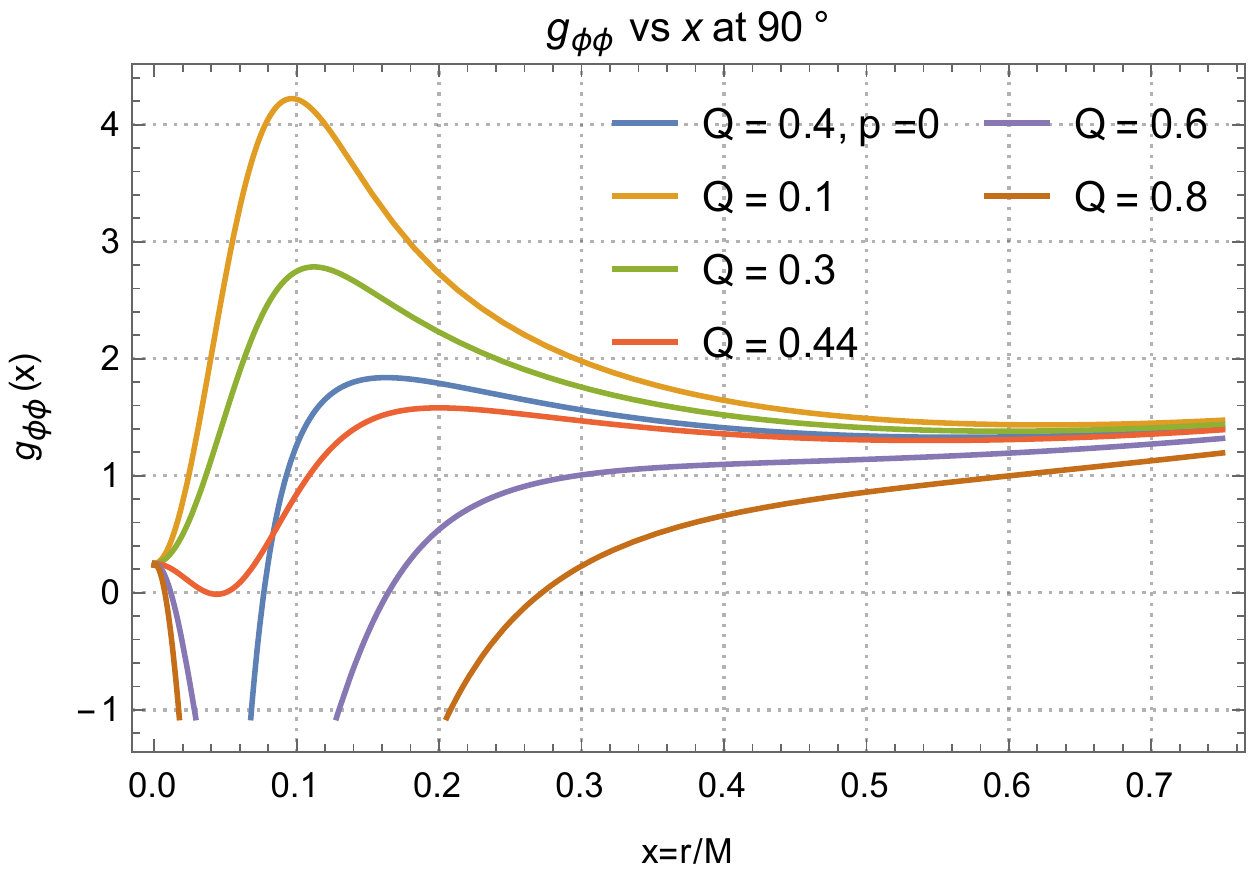}
\includegraphics[width=0.49\textwidth]{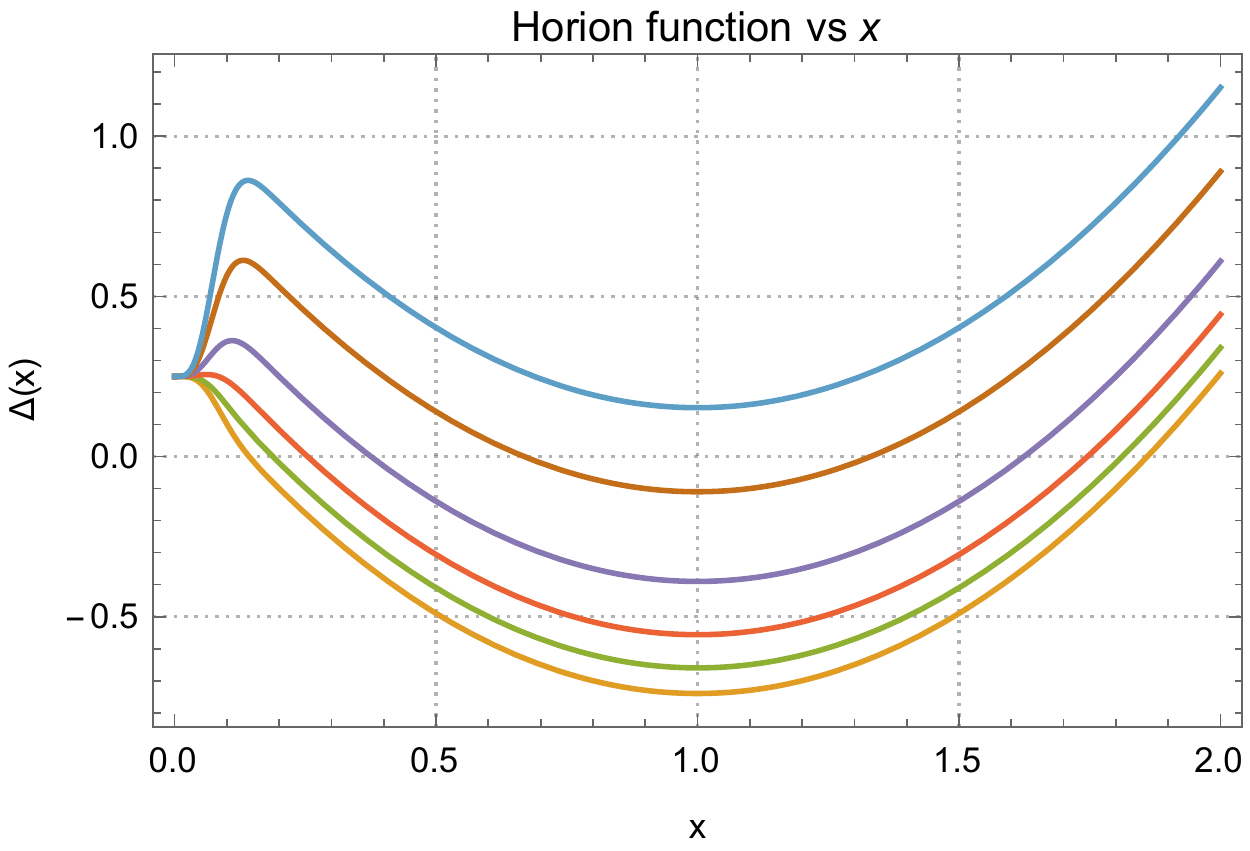}
\caption{Plots of the metric component $g_{\phi\phi}$ vs. $r/M$ for $a'= 0.5$, $p = 10^{-3}$. It is seen that for q' around $0.44$, the CTC region vanishes for non-zero p. The usual Kerr-Newman case ($p \rightarrow 0$) is depicted in dark blue for comparision. Note that the curve diverges as $r \rightarrow 0$ . Accompanying panel on the right shows the horizon function $\Delta(x)$  for the same values of BH parameters on the left. }
\label{fig:NCKNgppxQ}
\end{figure}

In this section, we will explicitly show that CTCs also exist in the Non-Commutative KN BH. We will then discern the grometrical properties of the CTC region and analyse geodesics that can visit the acausal region.
  
In order to find out whether or not a region containing CTCs exist in the NCKN spacetime, we need to analyse the $g_{\phi\phi}$ component of the metric:

\begin{align}
g_{\phi\phi} = sin^2\theta\left(\rho^2 +a^2sin^2(\theta)\left(2-\dfrac{\Delta-a^2 sin^2\theta}{\rho^2}\right)\right)
\end{align}

Where the symbols appearing in the above equations are as defined above in \eqref{NCKNsymb}. 

In particular, CTCs exist if the above function is allowed to take negative values in certain regions of the manifold. For this, we will begin by looking at the plots of the metric component $g_{\phi\phi}$ on the equatorial plane. It will be sufficient to understand the geometrical structure part of the CTC region lying on the equatorial plane. The structure of the CTC region boundary is very similar to that of the Kerr-Newman Black-hole as shown in the second figure Fig. \ref{fig:KNcont}.

Below we show a plot of $g_{\phi\phi}$ vs $r$ (measured in mass units) for different values of the BH parameters $\{a, q, p\}$ on the equatorial plane $\theta = \pi/2$.  In the first plot ${a = 0.5,q = 0.25,p = 0.001}$ Fig. \ref{fig:NCKNgppxQ}, we see that the equation $g_{\phi\phi} = 0$\label{gpp0} has positive roots on the equatorial plane.  This implies the existence of a finite sized CTC region. The accompanying figure shows that these parameters correspond to a black-hole solution. Thus, CTCs are also admitted in the Non-Commutative Kerr-Newman spacetime. Now we explain a commonly held misconception regarding what seems to be a plausible solution to the problem of CTCs . This will also help understand clearly what is happenning in such manifolds that admit CTCs.

Consider the new vector fields defined by:

\begin{align}
& d\eta = dt - a sin^2\theta d\phi &  d\xi = \dfrac{(r^2 + a^2) d\phi - adt}{\rho^2}
\end{align}

Now consider a transformation of co-ordinates as given by the above equations. Then, the four vectors $\mathbf{e_{(\eta)}}^\mu,\mathbf{e_{(r)}}^{\mu} = (0,1,0,0), \mathbf{e_{(\theta)}}^{\mu} = (0,0,1,0), \mathbf{e_{(\xi)}}^{\mu}$ together form an orthogonal tetrad, in the basis of which the metric is diagonal with the components:

\begin{align*}
g_{(\mu\nu)} = 
\begin{bmatrix}
    -\dfrac{\Delta}{\rho^2}    & 0                         & 0     & 0 \\
    0                          & \dfrac{\rho^2}{\Delta}    & 0     & 0 \\
    0						  &0 						  &\rho^2 & 0 \\	
    0                          & 0                         & 0     & \rho^2 sin^2\theta
\end{bmatrix}
\end{align*}

Therefore in this basis the vector field $e_{(\eta)}^\mu$ is time-like outside the outer horizon and inside the inner horizon, and space-like in between.(Note that indices with a bracket are tertrad indices).

It is evident that, by formally setting $a\rightarrow 0$ we recover the usual Kerr-Schild time and azimuthal co-ordinates $t$ and $\phi$. Thus, these new vector fields can be called the new "time" and "angular (azimuthal)" co-ordinates of the transformed co-ordinate system, which replace their counterparts in the non-rotating scenario ( i.e. Schwarzschild / Reissner-Nordstrom solutions). 

It can now be seen that the metric component $g_{(\xi\xi)}$ in the new basis is strictly positive. Therefeore it appears as though there are no CTCs in the new co-ordinate system, as the new azimuthal 'angle' is always spacelike. However, this is not the case. It is to be noted that , while the topology of the integral curves of $e_{(t)}^\mu$ and $e_{(\phi)}^\mu$ in the non-rotating scenario  were $\mathbb{R}$ and $S^1$ respectively, the integral curves of $e_{(\eta)}^\mu$ and $e_{(\xi)}^\mu$ is now $\mathbb{R}$ for both for finite $\frac{dt}{d\phi}$. This is due to the "mixing" of co-ordinates ($t$ and $ \phi$) due to dragging/rotation of the spacetime. They are now both helical (except at the poles). What we have made is just a co-ordinate transformation equipping the same manifold with a different set of co-ordinates.  However, this does not mean that the closed time-like curves do not exist nor do they disprove their existence. In doing the trasformation, the topological properties of the manifold remain unchanged. We then understand that the existence of closed time-like curves, although true, is merely not evident in the new co-ordinate basis. 

We now analyse and describe the dependence of the geometry of CTC region on the Black hole parameters. In the NCKN black hole, the metric is a function of three paramaetrs: the specific angular momentum and charge, and the non-commutativity parameter which we have denoted by $p$. The metric component $g_{\phi\phi}$ is also a function of these three parameters and thus the CTC region can be analysed with respect to them.

For a given set of values of parameters ${a,q}$, one can vary the non-commutativity parameter $p$ and understand the variability of the CTC region with respect it. Fig.\ref{fig:NCKNgppxp} shows the same. The first major effect of non-commutativity is that $g_{\phi\phi}$ remains finite as $r \rightarrow 0$. Secondly, we see that the metric component takes larger negative values and the equatorial size of the CTC region is larger ( i.e. separation between the two roots is larger) for smaller p. Thirldy, for a given set of parameters {$a,q$}, there exists a $p = p_{cr}$ for which the entire CTC region degenerates into its boundary, and the region disappears altogether for higher values of $p$. For instance, for the case of {$a=0.5,q=0.5$}, the two positive roots merge at $p$ (approximately) around $0.16$. There are no roots to the equation \eqref{gpp0} for a larger $p$. The adjoining panel shows the profile of the horizon function with $x$. They show that states corresponding to these values of $p$ do correspond to black holes with two horizons. There is no significant effect of variation (or introduction) of p on the placement of the horizons as expected. It can be seen that there is only some variation on the scale of $\sqrt{p}$, which here is much smaller than the horion scale. 
Thus, there exists a critical value of $p_{cr}$ for given set of values of the parameters $a,q,\theta$ such that the CTC region exists. For all values of $p$ above this value (super critical), the CTC region vanishes. i.e.,
\begin{align}
p<p_{cr} \qquad \textit{CTC region exists}\\
p>p_{cr} \qquad \textit{No CTC region}
\end{align}
This is an importnant feature of the Non-commutative black hole model.
It is interesting to investigate the energetics of this process in the above two cases. As non-commutativity introduces a minimal length scale, it will be interesting to understand if $p_{cr}$ can be associated with the Planck-scale $M_{pl}$.  Note that the critical value  $p_{cr}$ depends on the other parameters.

The observations made above can be reasoned out as follows:  The Kerr ST has no CTC region due to the absence of charge. The non-commutativity in the NCKN black-hole effectively replaces the ring singularity with a distribution (of mass and charge). Hence, due to this spread,  at small r, the Kerr character is more dominant than the KN character in the space-time. The Kerr-Newman behavior is expected to dominate at large distance from the center i.e. at distances larger than a few times the non-commutative length-scale $\sqrt{p} $.

Also, since the mass $M$ also has a Gaussian spread that is flat in the neighborhood $r= 0$, the metric is expected to resemble the flat space-time with a constant positive energy density i.e. the de-Sitter metric.

The next set of figures \ref{fig:NCKNgppxQ} show that for a BH woth given spin parameter $a/M$ and non-commutative parameter $p$, the region containing closed time-like curves exists in only those BHs with sufficiently high specific charge $q$. The CTC region does not exist for $q$ smaller than a limiting value which will be a function of the other BH parameters. The adjoining figure shows that BHs correponding to each of these values admit two horizons. However, for large enough q, horizons can disappear.

From the next figure Fig. \ref{fig:NCKNoutRoaQ}, we can infer that the outer boundary of CTC region does not monotonically increase with the specific charge $a'$, as expected (as in the case of Kerr-Newman spacetime).
So for a given a set of parameters of the blackhole: $\{a, q\}$, the above arguments indicate that for subcritical $p$ ($p<p_{cr}$), the inner horizon continues to be a Cauchy horizon even though there is no singularity in the spacetime. This is purely due to the pathology of the causality violating region which still exists. The inner horizon continues to be the Cauchy Horizon and the region within the inner horizon is classified as a 'viscious set' since the region is in causal contact with a causality violating region. For  super critical $p> p_{cr}$, the CTCs dissappear and therefore Chronology is protected.  The inner horizon is no longer the Cauchy horizon since the singularity has been smeared out and also the causality violating region eliminated. 


Thus, although the Non-Commutative Kerr-Newman space-time does not have curvature singularities , the closed time-like curves nevertheless still exist. This clearly shows that the problem of causality violation is not completely cured in Non-commutative space-time but it can be circumvented by choosing the parameter $p$  such that Chonology is protected.

\begin{figure*}[htbp]
\begin{minipage}{0.5\textwidth}
\centering
\includegraphics[width=\textwidth]{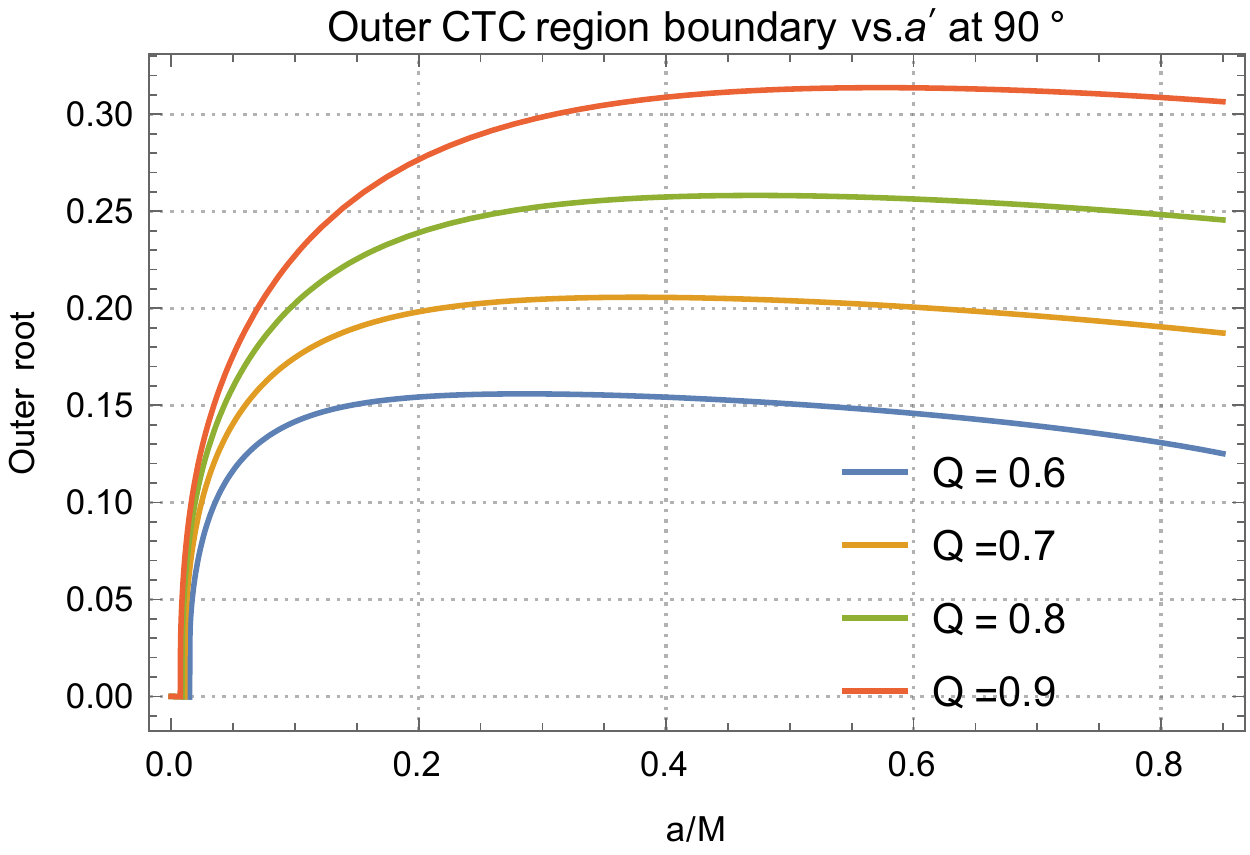}
\end{minipage}
\caption{A plot of the position of the boundary of the CTC region vs. the specific angular momentum a'. The boundary extends farther out for higher q' as expected.}
\label{fig:NCKNoutRoaQ}
\end{figure*}

\subsection{\label{sec:NCKNgeodesics}Causality violating geodesics}

Understanding the geodesic structure of a space-time is important in understanding how well the different regions of a particular space-time are connected. There exist significant literature on studies of geodesic structure of various black hole spaceitmes \citep{hackmann2013} \citep{saheb16}  spacetimes. Also, studies of those geodesics which can specifically carry causality violating data from the acausal regions of black holes can also be found in \citep{mthesis}. Given that CTCs also exist in the Non-commutative Kerr-Newman space-time, it is interesting to investigate geodesic connectivity of the CTC region of this spacetime. Specifically, we studied how exposed the CTC region is to the outer regions of the black hole and how different the geodesic structure is from the Kerr-Newman scenario. In this section, we present results of our analysis carried out using the Hamilton-Jacobi approach of the geodesic connectivity of the CTC region of the NCKN black hole. This will help us in understanding whether there exist any natural protection mechanisms in causality violating black holes that can preserve casuality in the region outside the black hole horizons. For achieving this, we primarily studied \emph{geodesic turning points} in the CTC region. Existence of turning points in the CTC region would enable geodesics from the outer regions to visit/sample points from the CTC region and turn back to reach the outer regions again, communicating causality violating data to them.
We begin by briefly illustrating the Hamilton-Jacobi approach in obtaining first integrals of motion and then using them in the analysis of turning points.

The action of a massive point particle of mass $m_0$ in a curved background is given by\footnote{There are other actions that lead to the same EOM}:
\begin{align}
\mathcal{A} = -m_0\int d\tau 
\end{align}
In the Hamilton-Jacobi theory, the action, which is a solution to the Hamilton-Jacobi (H-J) equation, is treated as a function of the co-ordinates and the components of momenta are given by:

\begin{align}
\nabla_\mu \mathcal{A}(x^\nu) &= p_\mu(x^\nu) \textit{or} \\
\end{align}
The normalization condition for time-like vectors leads to:
\begin{align}
g^{\mu\nu}\dfrac{\partial A_\mu}{\partial x_\mu} \dfrac{\partial A_\mu}{\partial x_\mu}  +m_0^2 = 0 \label{spnorm}
\end{align}

In the H-J theory, the symmetries of the space-time lead to constants of motion and facilitate a separation of variables of the action in the additive fashion. In this case, the background spacetime is the NCKN space-time and it can be checked that this space-time being stationary and axisymmetric like the usual KN space-time, admits a time-like Killing vector field $\xi^\mu_{(t)}$ (which we represent in the $(t,r,\theta,\phi)$ convention as $(1,0,0,0)$), an axisymmetric Killing vector field $\xi^\mu_{(\phi)} = (0,0,0,1)$ and also a Killing tensor field $K^{\mu\nu}$ analogous to the KN scenario (see \citep{carter1968}). The Killing vectors facilitate the separation of the time and the azhimuthal angle co-ordinates while the Killing tensor is responsible for fully separating the action by separating the $\theta$ angular co-ordinate from the radial co-ordinate $r$.  The symmetries also lead to conserved quantities, i.e. quantities that are constants on each geodesic and can be identified with certain physical quantities. These are given as:

\begin{align}
E = -\mathbf{p}\cdot\mathbf{\xi_{(t)}} & & L = \mathbf{p}\cdot\mathbf{\xi_{\phi}}
\end{align}

The first two can be readily identified with the total energy and azimuthal angular momentum of the particle whereas $Q_0 ^2$ is known as Carter's constant introduced first by Carter in \citep{carter1968}.

Thus fully separable action can now be written as:  

\begin{align}
\mathcal{A} = -Et + L\phi + \int p_r(r) dr + \int p_\theta (\theta)
\end{align}

This action when substituted in \eqref{spnorm} using the inverse metric components leads to the following expressions for the radial momentum $p_{r}$:

\begin{align}
p_r^2 (r) = \dfrac{1}{\Delta^2}\left( - \Delta(aE-L)^2 + ((a^2+r^2)E - al)^2 -(Q_0^2 +m_0^2)\Delta\right) \label{NCKNpr}
\end{align}

Firstly, we can readily see that in the limit $r \rightarrow 0 $, the RHS of the equation becomes negative due to non-zero $(m_0^2)$  and $( Q_0^2)$ terms. Since the particle cannot exist in the regions where $p_r^2 <0$, one can conclude that no time-like particles can reach the region $r = 0$ (first shown by Carter \citep{carter1968}).They turn back before that. However, it can be verified that special null curves on the equatorial plane satisfying the condition $Ea - L$ can reach the ring singularity. The significance of this condition will be explained later in Sec. \ref{sec:disc}. Thus, like in the KN case, $r=0$ is accessible to only null geodesics restricted to the equatorial plane. These geodesics can carry both the data from the causality violating region and $r=0$.

Secondly, it is to be noted that the existence of a turning point for null geodesics in the CTC region is a sufficient condition to have them carry causality violating data to the outer regions of space-time. This is true because we cannot have turning points in between the horizons. The corresponding equation for null-geodesics can be obtained by formally setting $m_0^2 \rightarrow 0$ in the above equation.

We now demonstrate the existence of geodesics that can communicate causality violating data to the outer regions of the BH. There are specifically two cases to consider: One for which the BH has no horizons, which would have been called a naked singularity but now is regular, and second for the case with horizons. We will discuss the result in the context of these two cases towards the end.

\begin{figure*}[htbp]
\includegraphics[width=0.49\textwidth]{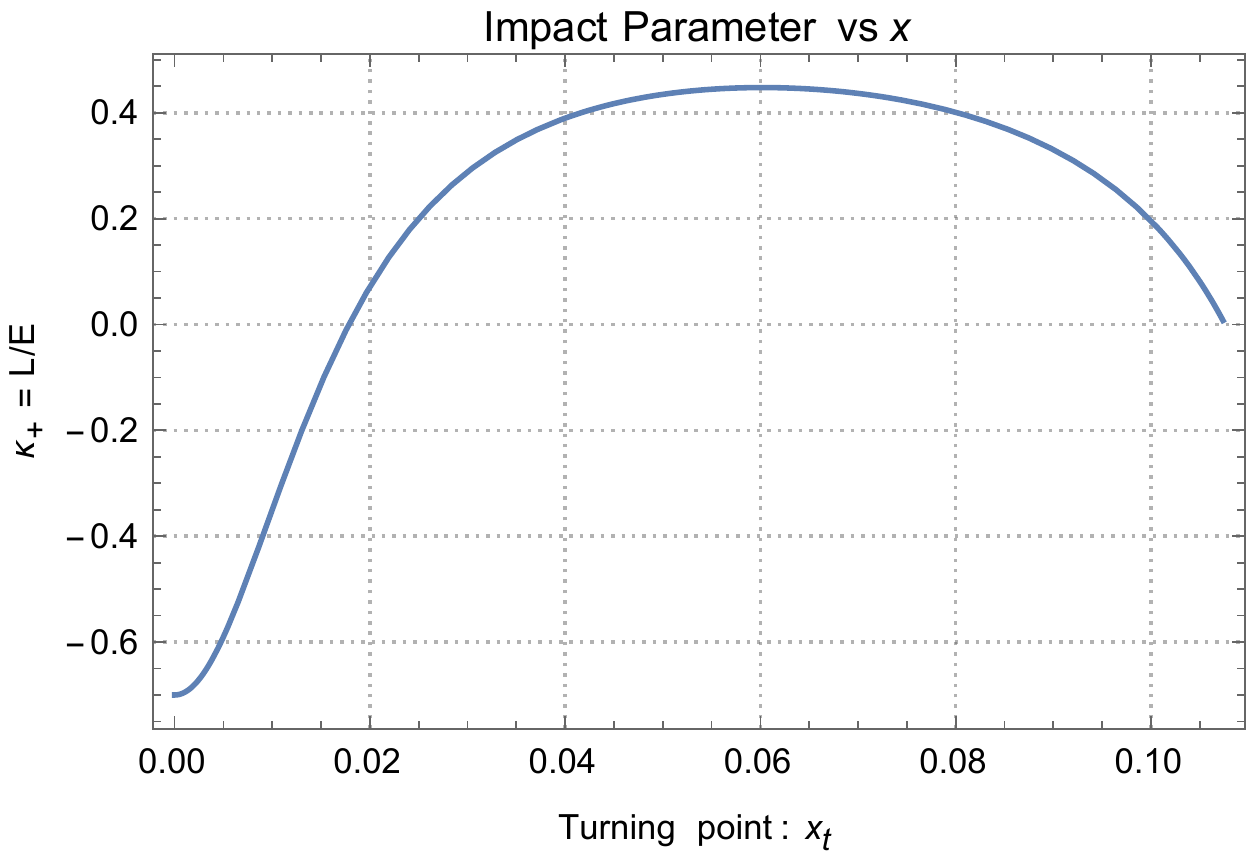}
\includegraphics[width=0.49\textwidth]{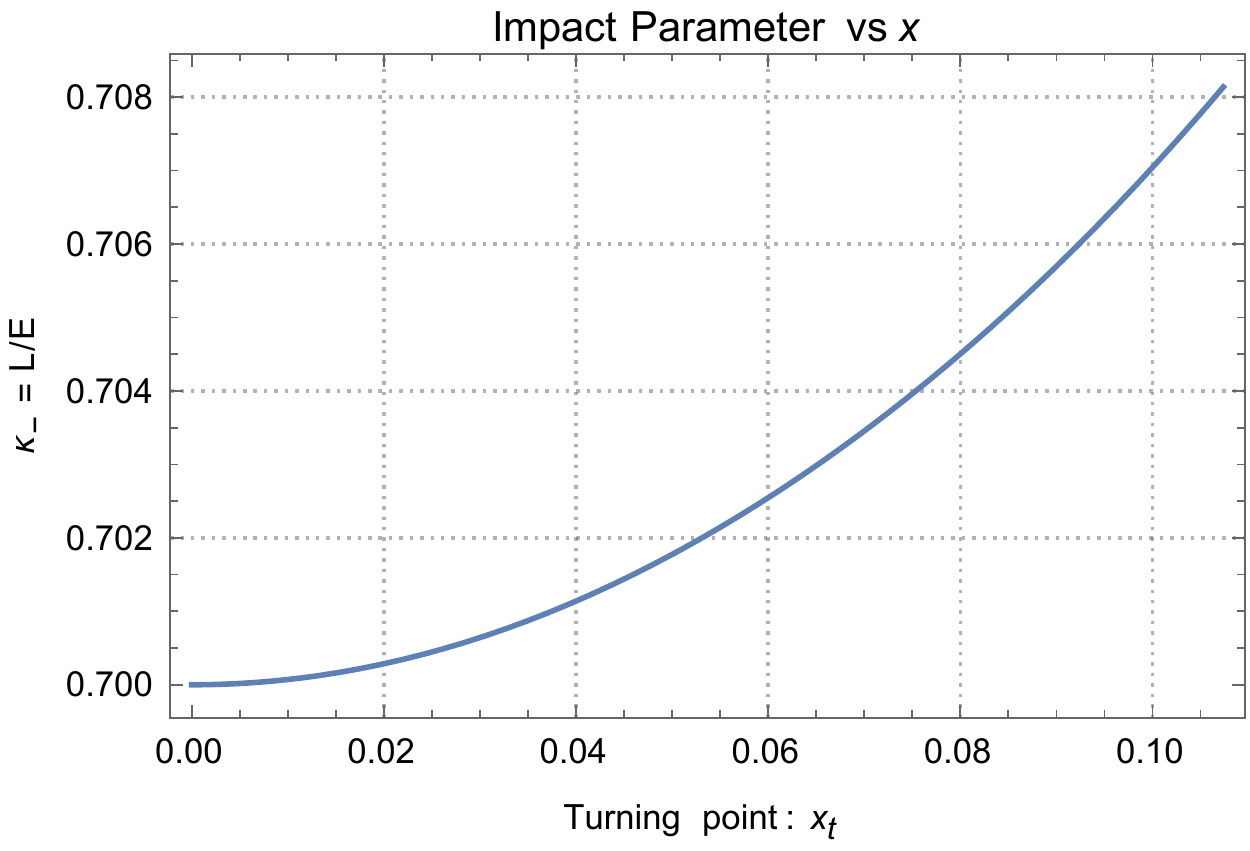}
\caption{Plot of the geodesic impact parameter $\kappa = L/E$ for which the geodesic has a turning point at $x_t$. The domain of $x_t$ has been chosen so as to coinscide with the region containing CTCs.}
\label{fig:NCKNk}
\end{figure*}

For the case of a null geodesic, it is easy to demonstrate the existence of such a geodesic. Consider the above radial equation on the equatorial plane $\theta = \pi/2$. Using $g_{rr}p^r = p_r$ and $p^r = \frac{dr}{d\lambda}$ for an affinely parametized null geodesic with affine parameter $\lambda$, we can rewrite the above equation as:
\begin{align}
\dfrac{d r}{d\lambda}^2 = \dfrac{1}{r^4} \left( -\Delta (aE - L)^2 + ((a^2 + r^2)E-aL)^2\right)
\end{align}

If we choose the constants of motion such that they follow the relation $Ea=L$, this geodesic equation becomes $dr/d\lambda=E^2$. The motion is therefore unbounded. A ray with the above set of parameters can emanate from the CTC region and reach outside points. Also, rays starting from the asymptotically flat region can travel inwards, cross the CTC region and reach $r=0$. Since $r=0$ is not a singular point, the geodesic can be continued to arbitrary values of the affine parameter.  The pathology carried by the null ray comes from it's passage through the CTC region. 

For different sets of parameters, we now look for the turning points of the null-curve on the equatorial plane. A turning point is given by $dr/d\lambda=0$. This yields 

\begin{equation}
 (r^2 + a^2)E - aL)^2 - \Delta(aE-L)^2 = 0\\  
\end{equation}

A condition on the parameters $\{E,L\}$ for null-geodesics to have a turning point in the CTC region can be derived by rearranging and treating this as a quadratic equation in $L/E$, which we denote by $\kappa$:

\begin{align}
(a^2 - \Delta) \kappa^2 + 2a(\Delta - (a^2+r^2)) \kappa + (a^2 +r^2)^2 - a^2\Delta = 0 
\end{align}

The above equation has the solutions:

\begin{align}
\kappa = \dfrac{a(a^2 +r^2 - \Delta) \pm r^2\sqrt{\Delta}}{(a^2 - \Delta)}
\end{align}

This equation summarizes the results we have found so far succinctly:
\begin{itemize}
\item There can be no turning points in between the horizons, as one expects, since $\Delta < 0 $ there.
\item If the turning point is at $r \rightarrow 0$, then $Ea = L$.
\item For all other radii in the CTC region i.e. $r_{c_-} < r < r_{c_+}$, $L/E$ takes the values shown in the figure \ref{fig:NCKNk}
\end{itemize}

It must be remarked that this equation does not take into account any constraint on $ L/E$ that comes from basic dynamics. In Fig. \ref{fig:NCKNk}, we plot the parameter $\kappa = L/E$ of the geodesic that has a turning point at $ x_t$. The domain of the plot has been chosen such that it contains the CTC region.

\subsection{\label{sec:NCKNconsequences}Consequences}

Now we discuss this result in the context of the two cases mentioned above:
\begin{itemize}
\item For the case corresponding to no horizons
\item For the case corresponding to single or double horizons:
\end{itemize}

When the black hole has no horizons: Since a geodesic cannot have a turning point in between the horizons, it is then clear that a geodesic that visits the CTC region and turns back will carry causality violating data to outer regions of the spacetime. Thus, eventhough there exist no curvature singularities, the whole space-time is becomes a viscious set.

When the black-hole does have horizons, the inferences that can be drawn depends on whether or not one accepts maximal extensions of spacetimes. It is now well known that the co-ordinates on manifolds of black hole spacetimes (and some others) can be analytically continued onto other domains facilitating maximal extension of the space-time and ensuring geodesic completeness. From this point of view, we can conclude that causality violating data from the interior region corresponding to one asymtotically flat block can be communicated to another. However, if one allows the particle to emerge into the same region of asymptotically flat sapcetime from which it entered, then the causality violating data can be communicated to the outside observers.

In either of these cases, some of the asymptotically flat blocks will be viscious sets, apart from the inntermost region which is completely regular. 




\section{\label{sec:FRKNintro}Causality in f(R) Kerr-Newman Black-hole}
In this section, we focus on the causality aspects of the Kerr-Newman Black hole in the context of f(R) gravity as derived in \citep{cembranos14}. The Kerr Newman metric presented in the above reference is a solution to the vacuum field equations where it is assumed that the curvature scalar is constant $R_0$.
The objective is to find the causality violating regions in the modified Kerr-Newman metric and examine the possibility of eliminating the violating regions. Another objective is to examine if a suitable function $f(R)$ exists that can shrink the causality violating region to Planck scale so the region could be smeared out in a Quantum Gravitational setting. The work done here also encompasses the causality analysis of Kerr Newman De Sitter and Anti-De Sitter space-times since they form a subset of solutions of the more general framework presented in the above reference.
In this section we discuss the issue of violation of causality in f(R) inspired Kerr-Newman BH.

\subsection{\label{sec:FRKN}f(R) Kerr Newman black hole}

We now consider the Kerr-Newman solution of f(R) gravity given by the (incomplete) action \footnote{The boundary term is to be added to complete the gravitational action}:

\begin{align}
\mathcal{A} = \dfrac{1}{8\pi G} \int \sqrt{-g}d^4x (R +f(R))\label{FRact}
\end{align}

One of the main features of this theory is that it allows for non-zero scalar curvature solutions to the Field equations corresponding to the action \eqref{FRact} for the vacuum space-time. The condition $T_{\mu\nu} = 0$ leads to 
\begin{align}
R_{\mu\nu}(1+f'(R_0)) - \dfrac{1}{2}g_{\mu\nu}(R_0 + f(R_0)) = 0
\end{align}
i.e. the Ricci tensor of vacuum becomes proportional to the metric tensor ($for 1+f'(R)\neq 0$). Taking the trace, we find that 
\begin{align}
R_0 = \dfrac{2f(R_0)}{f'(R_0) - 1}
\end{align}

For the special case of constant scalar curvature $R_0$, the Kerr-family of the metrics, specifically the Kerr-Newman metric, have been derived \citep{cembranos14}. In Boyer-Lindquist co-ordinates, the metric of the Kerr-Newman space-time is given by,

\begin{align}
ds^2 = -& \frac{(\Delta_r - a^2\Delta_{\theta}sin^2\theta)}{\rho^2\Xi^2} dt^2 \\ -& \frac{2a sin^2\theta}{\rho^2\Xi^2}((r^2 +a^2)\Delta_\theta - \Delta_r)dtd\phi + \dfrac{\rho^2}{\Delta_r}dr^2 + \frac{\rho^2}{\Delta_{\theta}} d\theta^2\\ +& \dfrac{\Sigma^2sin^2\theta}{\rho^2\Xi^2 }d\phi^2 \label{frknmet}
\end{align}

with,

\begin{align}
\Delta_r = (r^2 + a^2)(1-\frac{R_0}{12}r^2) -2Mr+ \frac{Q^2}{1+f'(R_0)}\\
\Delta_{\theta}=1+\frac{R_0}{12}a^2cos^2\theta \qquad \rho^2 = r^2 + a^2cos^2\theta\\ 
\Xi=\:1+\frac{R_0}{12}a^2 \qquad \Sigma^2 =  \Delta_{\theta}(r^2+a^2)^2-a^2\Delta_r sin^2\theta
\end{align}

The new vector fields and the orthogonal tetrad specified for the NCKN black hole ( see \ref{sec:NCKNCTC}) are now modified into

\begin{align}
& d\eta = \dfrac{dt}{\Xi} - \dfrac{a sin^2\theta d\phi}{\Xi} &  d\xi = \dfrac{(r^2 + a^2) d\phi - adt}{\Xi\rho^2}
\end{align}

\begin{align}
g_{(\mu\nu)} = 
\begin{bmatrix}
    -\dfrac{\Delta}{\rho^2}    & 0                         & 0     & 0 \\
    0                          & \dfrac{\rho^2}{\Delta}    & 0     & 0 \\
    0						  &0 						  &\dfrac{\rho^2}{\Delta_\theta} & 0 \\	
    0                          & 0                         & 0     & \Delta_\theta\rho^2 sin^2\theta
\end{bmatrix}
\end{align}

The rest of the arguments follow as in the previous section.

\begin{figure*}[htbp]
\includegraphics[width=0.49\textwidth]{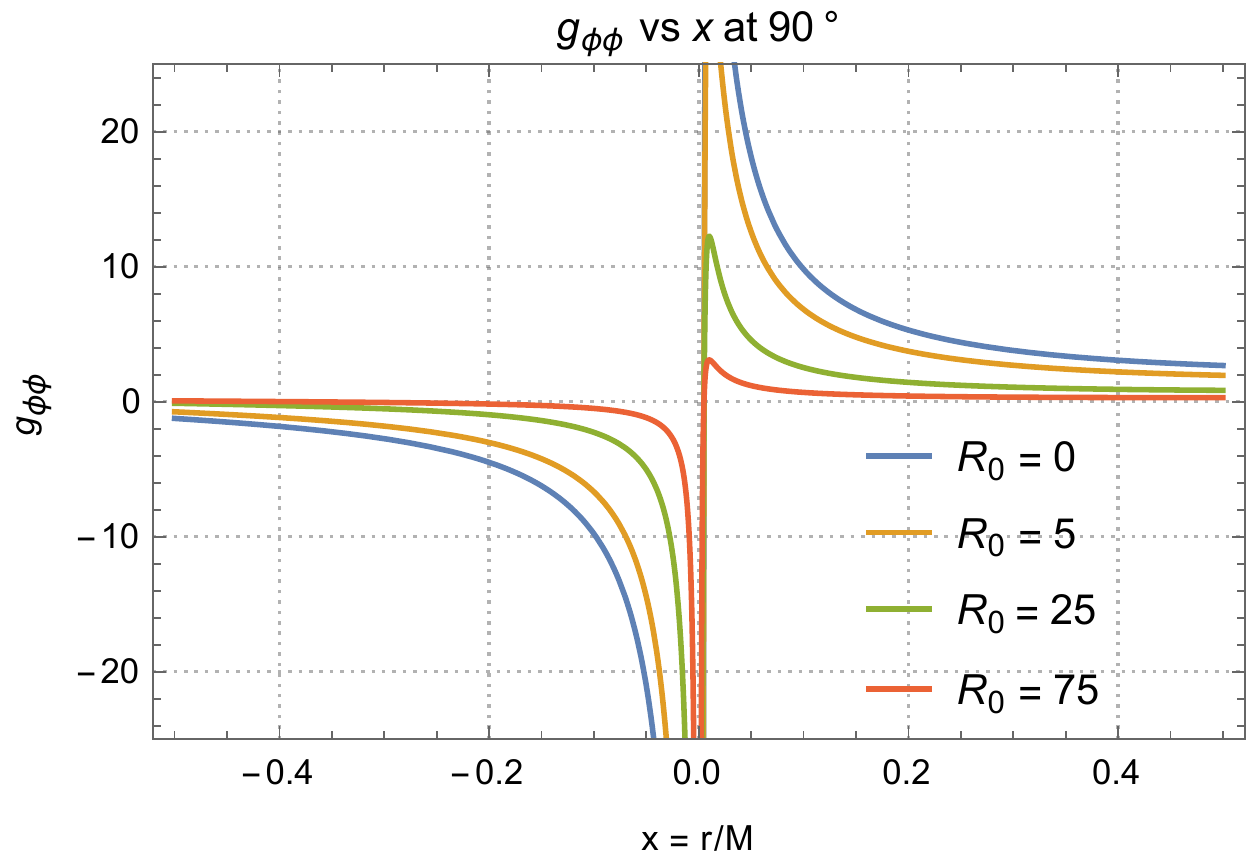}
\includegraphics[width=0.49\textwidth]{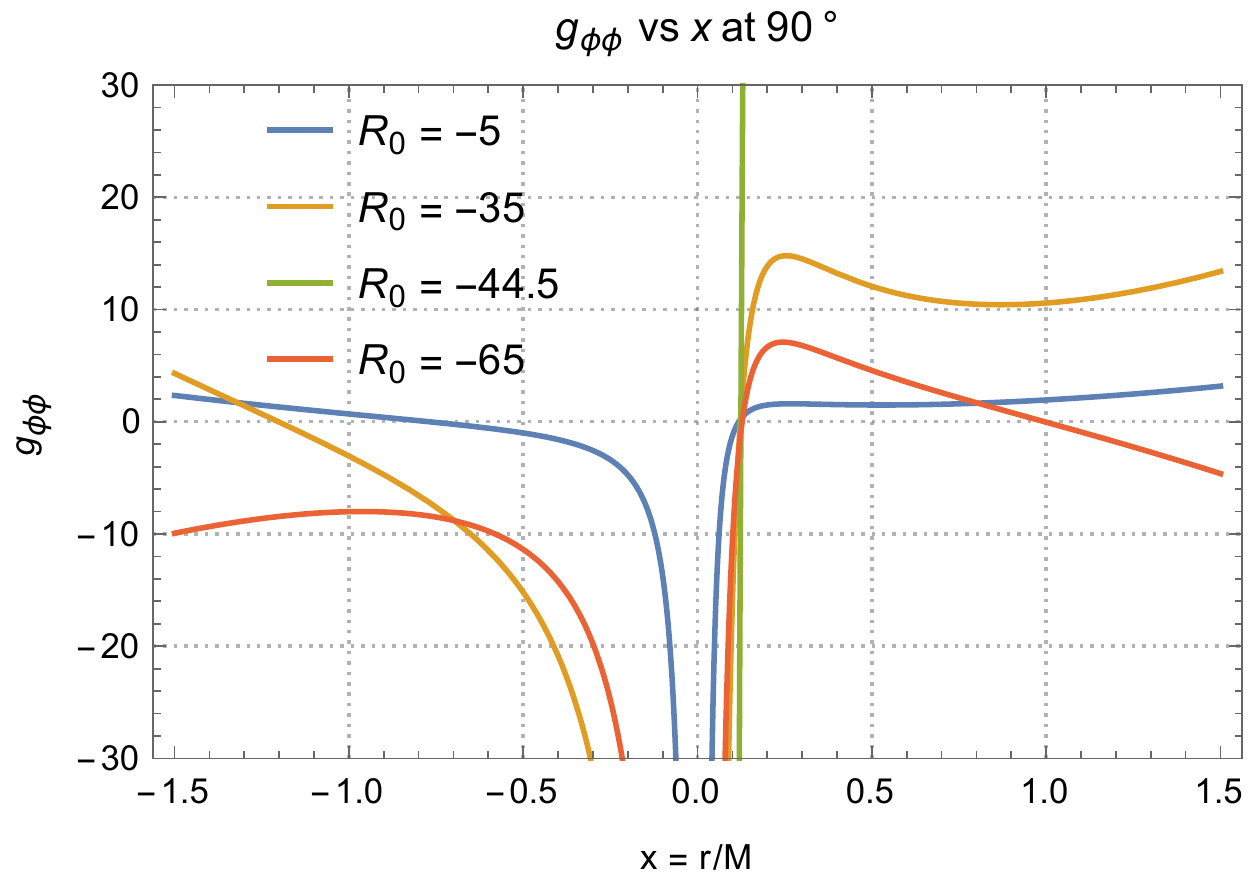}
\caption{The left curve shows the plot of $g_{\phi\phi}$ vs. x for different positive $R_0$, and on the right for different negative $R_0$. For sufficiently negative $R_0$, it can be seen that $g_{\phi\phi}$ crosses the $x$ axes again at a later $x = x_a$. Thus the region with $x> x_a$ also contains CTCs. This includes the asymptotic infinity region.}
\label{fig:FRKNgppxR}
\end{figure*}

The structure of the metric is similar to the usual KN metric \eqref{KNmetric} of \ref{sec:CPC}. A brief summary of the main features of the f(R) Kerr-Newman black hole given by the above metric is as follows:

\begin{itemize}
\item The number of horizons posse f(R) KN space-time can be anywhere between zero to three. This includes both black hole horizons (the Event horizon and the Apparent) and Cosmological Horizons. When three horizons exist, two of them can be interpreted as BH horizons and the outer-most horizon will be a Cosmological Horizon. Apart from this case, there can exist situations in which some of these horizons are degenerate ( or extremal) or there also exist no horizons i.e. a naked singularity. A detailed discussion can be found in \citep{cembranos14}

\item There are constraints on the modified Lagrangian $f(R)$ from cosmology and General Relativity. These are that the quantity $(1+f'(R))$ is always greater than zero and $f'(R)<0$ ( where dash represents derivative with $R$). This ensures that the effective gravitational constant is positive and the theory is free of ghosts \citep{cembranos14}. 

This allows us to define an effective electric charge:

\begin{align}
q^2 = Q^2/(1+f'(R)) \label{FRnorq}
\end{align}
which casts the metric in a form that resembles the usual Kerr-Newman metric \eqref{KNmetric} more closely. This allows us to examine the geometry of the CTC region in a model independent manner i.e. without having to choose a particular function $f(R)$ ( it more convenient to understand the CTC region of the f(R) Kerr-Newman space-time) . Note that this implies $q\geq Q$. The parameters that now need to be used to analyze the potential causality violating region of the Kerr-Newman space-time will be the dimensionless counterparts of the BH parameters {$a' = a/M, q' = q/M$} and the constant vacuum Ricci scalar, $R'_0 = R_0 M^2$. 

\item The Kerr-Newman De Sitter/Anti De Sitter metric of Einstein's theory is a special case of the $f(R)$ theory of gravity with $f(R) = -2\Lambda, R_0 = 4\Lambda$. Thus by analyzing Causality violation in the f(R) Kerr-Newman BH, the analysis of Causality in  Kerr-Newman dS / AdS would also have been achieved. We will discuss the implication of our results for this space-time in the later section of this paper.
\end{itemize}

\subsection{\label{sec:FRKNC}Causality violation in f(R) KN BH}

In order to discuss the validity of causality, we will again have to analyse the azimuthal component of the metric $g_{\phi\phi}$ of the Kerr-Newman black hole in the f(R) model of gravity:

\begin{align}
g_{\phi\phi} = \dfrac{\Sigma^2sin^2\theta}{\rho^2\Xi^2 }
\end{align}

Proceeding along the lines of the analysis done in the previous section, we will now discuss the corresponding causality violating region.  In contrast to the Non-commutative models, we have one new parameter with respect to which the behaviour of $g_\phi\phi$ can be discussed: the vacuum Ricci scalar $R_0$. We will begin therefore by analyzing the effect of $R_0$ on the CTC region. Specifically we will discuss the two dimensional region containing closed time-like curves on the $\theta = pi/2$ surface, the equatorial plane.   On the left top panel, we show a plot of the metric component $g_\phi\phi$  vs $r$ (normalised, on the equatorial plane), for different values of the vacuum scalar curvature $R_0$ . Some features to be notices are: 1. For positive values of $R_0$, the CTC region on the equatorial plane is suppressed in size with increasing $R_0$ ( i.e. $r_{ctc_-}$ and $r_{ctc_+}$ approach each other) and can be made to disappear for very large positive values of it.  Note that we cannot plot the corresponding figues of the horizon function $\Delta_r$ to discern the number of horizons for these parameters. This requires choosing a particular $f(R)$ model.  2. The CTC region on the equatorial plane becomes more prominent as $R_0$ takes larger negative values. 

As the metric components take a form similar to the Kerr-Newman scenario, with the metric characterised by the other parameters viz. the mass, specific charge ( effective, as defined above \eqref{FRnorq}) and angular momentum, the dependence of the properties of the CTC region may be expected to be similar to the Kerr-Newman case.
Note that we have not assumed any model in order to obtain these plots, which only reflects in the freedom in the choice of upper bound on $q$. Thus these results hold generally for any $f(R)$ model. 

\begin{figure*}[htbp]
\includegraphics[width=0.49\textwidth]{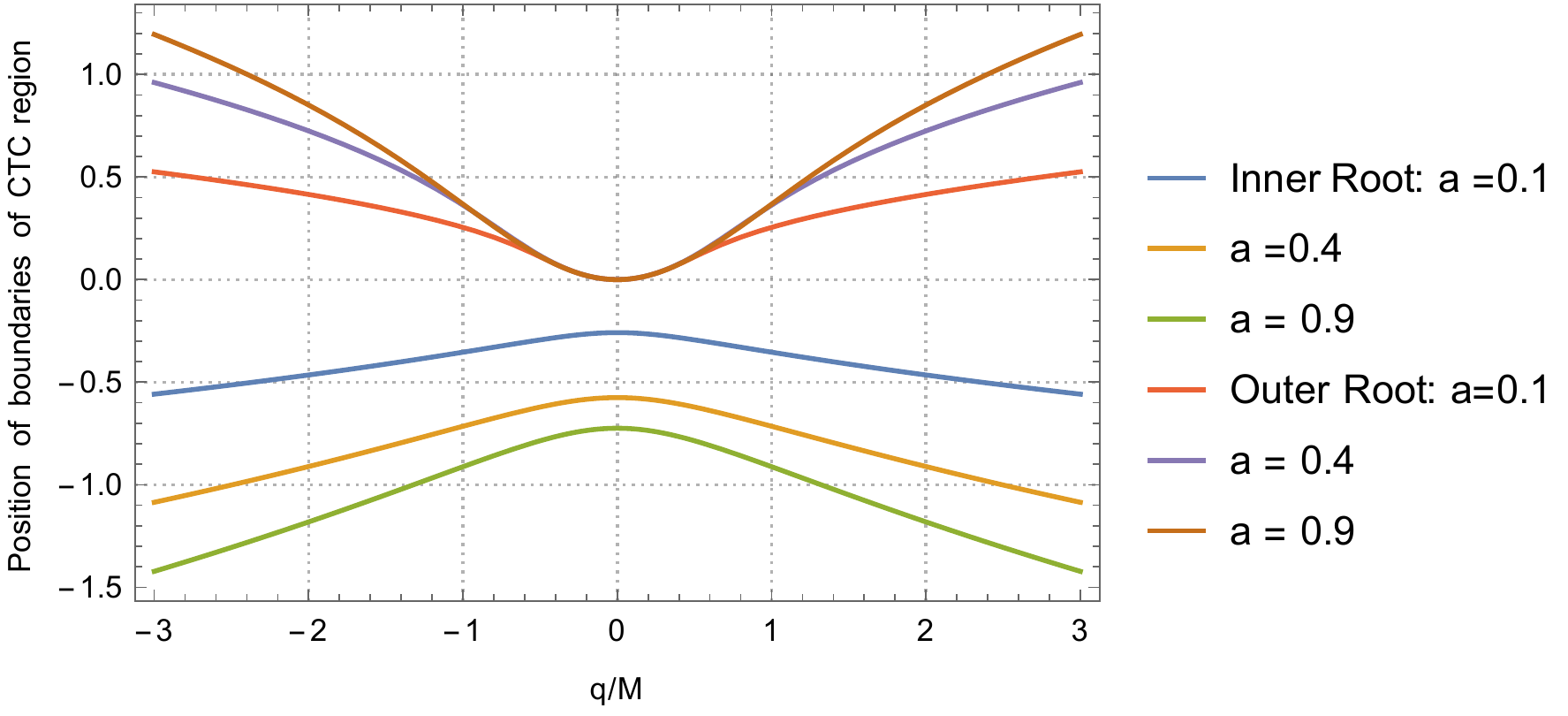}
\includegraphics[width=0.35\textwidth]{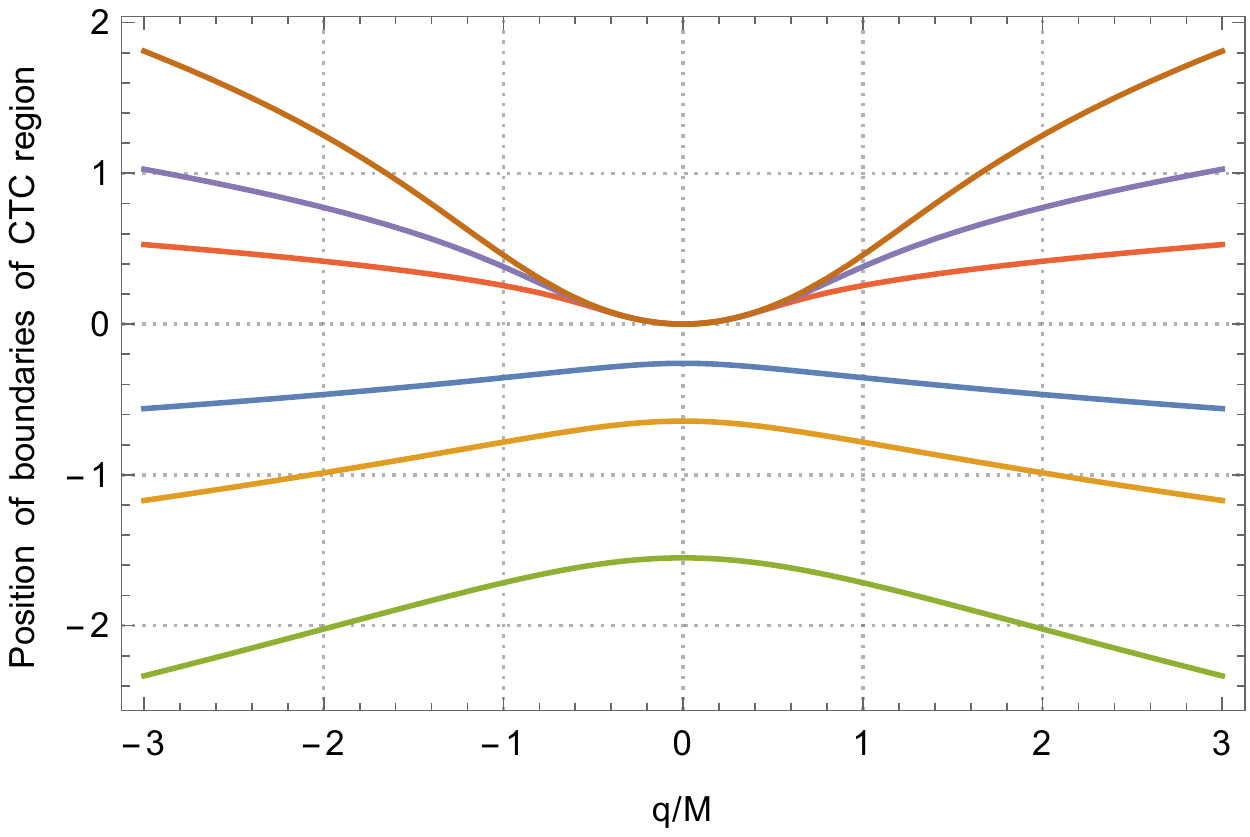}
\caption{Variation of the outer and inner roots of the CTC region with q. The left panel shows the plot for $R_0 = 10$ and the right for $R_0 = -10$. It can be seen that the CTC region increases in size on the equatorial plane in both the cases with q. The curves have been shown for different $a$.}
\label{fig:FRKNRoaq}
\end{figure*}


The main inferences that can be drawn from the graphs are that, the modification of the geometry of the spacetime due to $f(R)$ corrections do not change the qualitative nature of the curves

\begin{figure*}[htbp]
\includegraphics[width=0.49\textwidth]{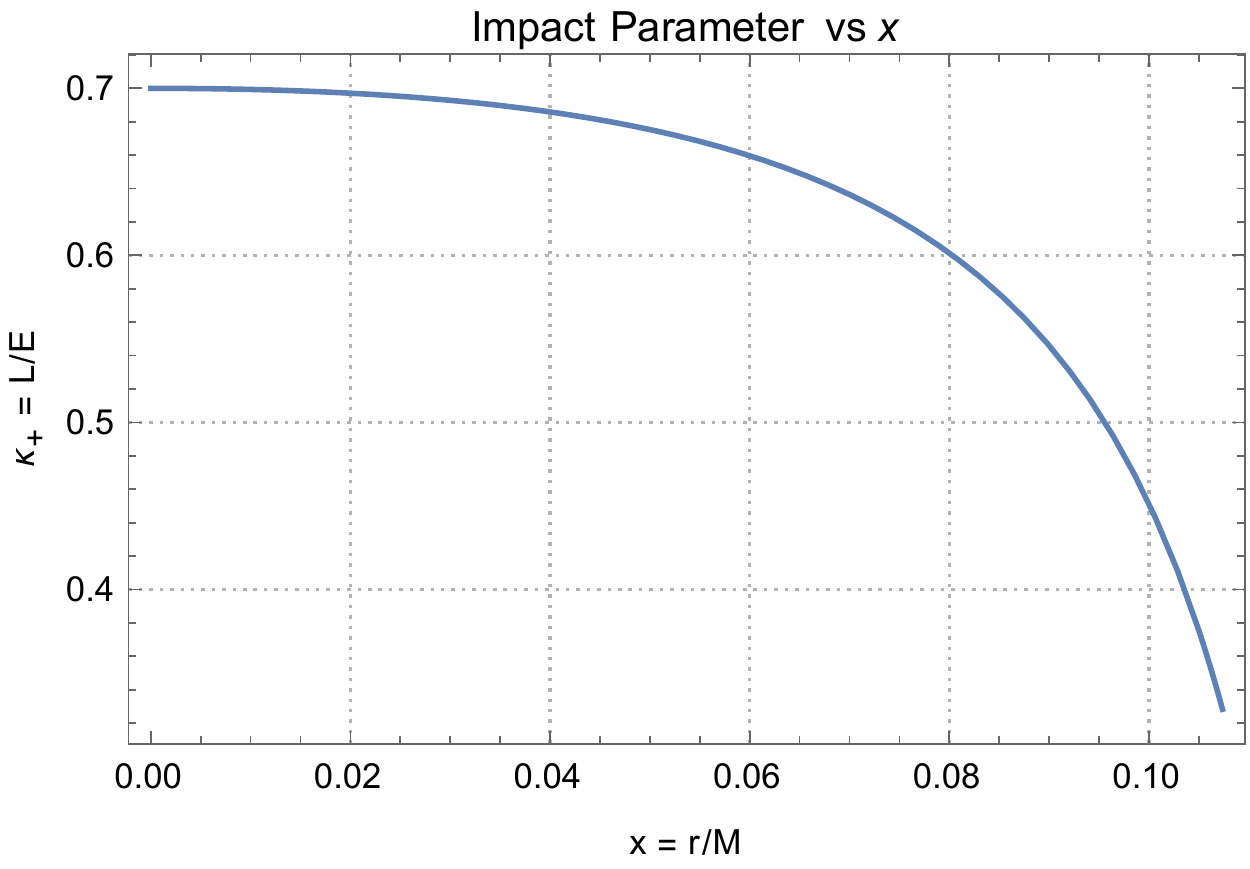}
\includegraphics[width=0.49\textwidth]{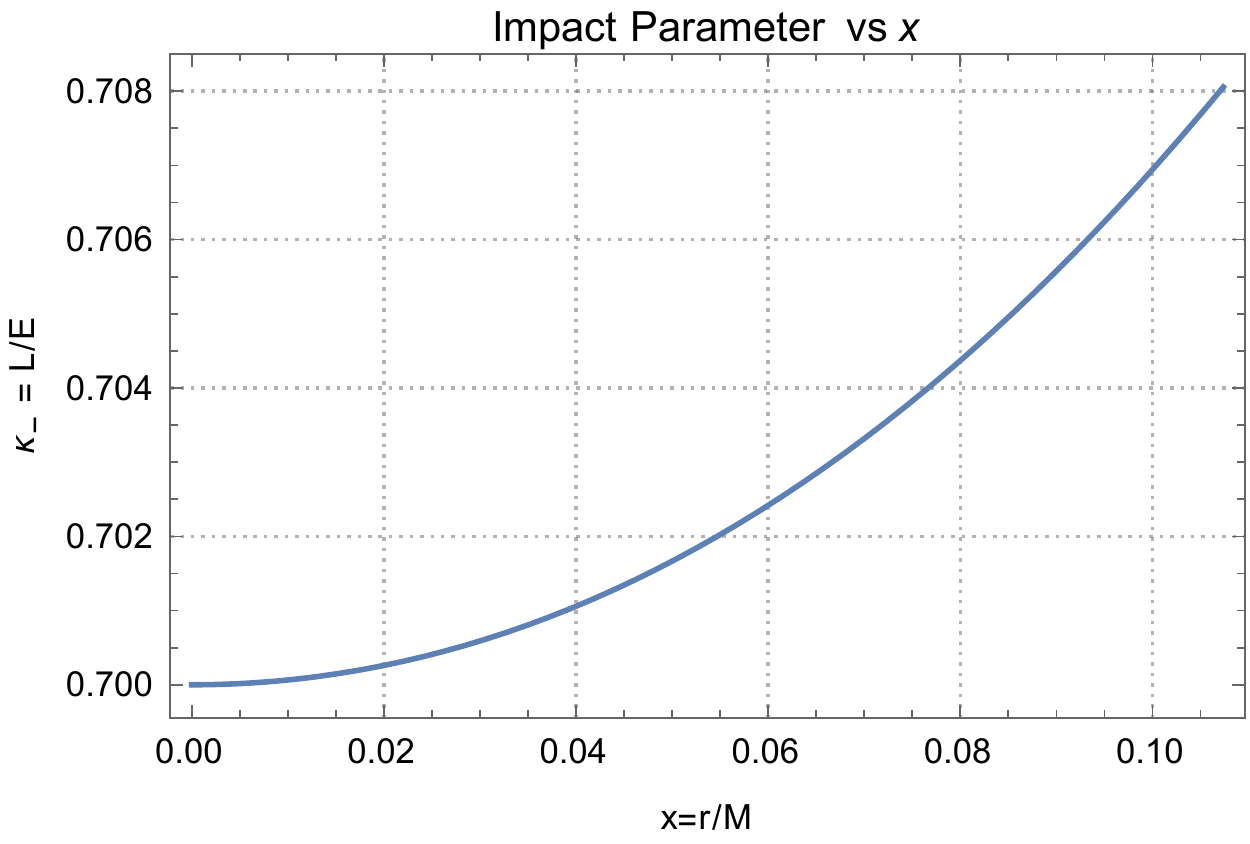}
\caption{Plot of $\kappa = L/(\Xi E))$ for the f(R) KN BH vs. the turning point radius on the equatorial plane. The domian of $x_t$ has been chosen so as to overlap with the CTC region.}
\label{fig:FRKNk}
\end{figure*}

\subsection{\label{sec:FRKNgeo}Causality violating geodesics}

Now, we consider the motion of null geodesics restricted to the equatorial plane in the Kerr-Newman black hole of f(R) gravity. The objective of the analysis again is to determine the existence of geodesics that can sample the region containing CTCs. As mentioned previously, for this purpose, it is sufficient to consider just equatorial geodesics.

Geodesics of the $f(R)$ gravity KN space time have been studied considerably. The geodesic structure of the sapcetime has been found to be quite similar to the KN scenario. An almost exhaustive clasiification of geodesics in the f(R) KN space time can be found in \citep{saheb16}. In the work, the behaviour of most of the geodesics has been detailed. In this section, we complete the description by additionally discussing \emph{causality violating geodesics}, the geodesics that can sample points from the acausal region. 

Following the same approach as in \ref{sec:NCKNgeodesics}, the radial equation for null geodesics can similary be obtained in the contex of the metric for f(R) Kerr-Newman BH 
(see \citep{saheb16}):

\begin{align}
\rho^4\left(\frac{dr}{d\tau}\right)^2=-\Delta(K)+(a\sqrt{K}+r^2E)^2
\label{frnull}
\end{align}

Similarly for time-like geodesics we have the following,

\begin{align}
\rho^4\left(\frac{dr}{d\tau}\right)^2=-\Delta(K+r^2)+(a\sqrt{K}+r^2E)^2
\label{frtimelike}
\end{align}

Here $K$ has the interpretation of the Carter constant, again arising from the existence of a Killing tensor . For the motion to be restricted to equatorial plane, the constant $K$ takes the value $K=aE-L\Xi$ as can be demonstrated from the $\theta$ equation in the reference (geodesic paper). $E$ and $L$ are the conserved quantities corresponding to the Killing vectors $\partial/\partial t$ and $\partial /\partial \phi$ respectively.

To identify the parameter range of geodesics that can have turning point at the radii where the closed time like curves are possible, we fix the radius $r=r_c$ where $r_c$ can be any point where $g_\phi\phi(r_c)<0$. we then solve the quadratic equation obtained from (\eqref{frnull}) and (\eqref{frtimelike}) by equating $dr/d\tau$ to zero.

We get the following equation for the null rays that have turning point at $ r= r_c/M$.

\begin{align}
(a^2-\Delta_{r_c})\kappa^2+2ar_c^2\kappa+r_c^4=0.
\end{align}

where $\kappa=a-(L/E)\Theta$ and $\Delta_{r}(r_c)$ is the function $\Delta_r$ evaluated at the turning point $r=r_c$. 

The solutions are symbolically similar to the NCKN case:

\begin{align}
\kappa = \dfrac{a(a^2 +r^2 - \Delta_r) \pm \Xi r^2\sqrt{\Delta_r}}{\Xi(a^2 - \Delta_r)}
\end{align}

A plot of $\kappa$ vs. $x_t$ is shown in the figures \ref{fig:FRKNk}. Again, the domain has been chosen so as to include turning points in the CTC region.

\subsection{\label{sec:FRKNconseq}Consequences}

Thus, as for the case of non-commutative black-holes, the black holes of f(R) gravity also possess closed time-like curves, that are unavoidable. The geometric structure of the CTC region is for the most part, similar to that of KN space time. The major features of the CTC region of the KN black hole model of $f(R)$ gravity are that for large negative values of the vacuum Ricci scalar $R_0$,   

The above results suggests that the CTC region can be made to shrink to as small a region as required by choosing a very $R_0$, which does not seem to be a very natural assumption.



\section{\label{sec:KNdS}The Kerr-Newman dS/AdS black hole}

In this sub section, we study the CTC region in the Kerr-Newman dS/AdS black hole. The Kerr-Newman dS/AdS space-time is a special case of the Kerr-Newman black hole in $f(R)$ Gravity theory with the identification (see \citep{saheb16}):

\begin{align}
R_0 = 4\Lambda & & f(R) = -2\Lambda
\end{align}

The metric obtained with the above substitution takes the same form as the general one \eqref{frknmet}, however with the simplification $f'(R) = 0$. As a result, the specific electric charge appearing in the metric components is the same as the redefined electric charge:
\begin{align}
q = \dfrac{Q}{(1+f'(R))^(1/2)} = Q 
\end{align}

The plot of $g_{\phi\phi}$ vs $x$ for the KN dS spacetime is similar to the one shown in Fig. \ref{fig:FRKNgppxR}. The the trend is similar to the previous plots as expected. The region containing closed time-like curves also exist in the Cosmological black-holes (Kerr-Newman dS/AdS). The CTC region extant on the equatorial plane decreases with increasing $\Lambda$ and increases with large negative values of it. Due to this behaviour, one may expect to get rid of the region containing CTC by assuming large values of the cosmological constant. However, this approach seems unreasonable as the value of cosmological constant required to avoid forming the CTC region is exorbitantly high. Alternatively, for a particular value of $\Lambda$, one can find an upper limit for the Mass of the blackhole such that the CTC region does not exist. This critical mass turns out to be extremely high compared to the masses of some of the largest known super massive blackholes. To illustrate this, assuming the Ricci scalar to only have contribution from the Cosmological constant($\Lambda = 1.11 \times 10^{-52} m^{-2}$, $R_0 = 4\Lambda$), consider the relation between the dimensionless Ricci scalar appearing in the equations above ( which we will denote here by $\chi$) and the mass of the black hole. Taking note that the actual Ricci scalar is related to the dimensionless one through $R_0 \rightarrow R_0 M^2$, we have:

\begin{align}
\dfrac{M}{M_\odot} = 1.28 \times 10^{23} \sqrt{\chi}
\end{align}

This suggests that, given the presently estimated value of the cosmological constant, the dimensionless Ricci scalar $\chi$ takes a very low value for even the most super-massive BH masses that we know of today. To exemplify this, consider the mass of the BH at the center of ( choose a galaxy), which is one of the most massive BH found yet. The value of $\chi$ for this mass turns out to be $ $. Thus, the prospect of shrinking the CTC region due to a high $R_0$ and getting rid of it is beyond hope.

\begin{figure*}[htbp]
\includegraphics[width=0.7\textwidth]{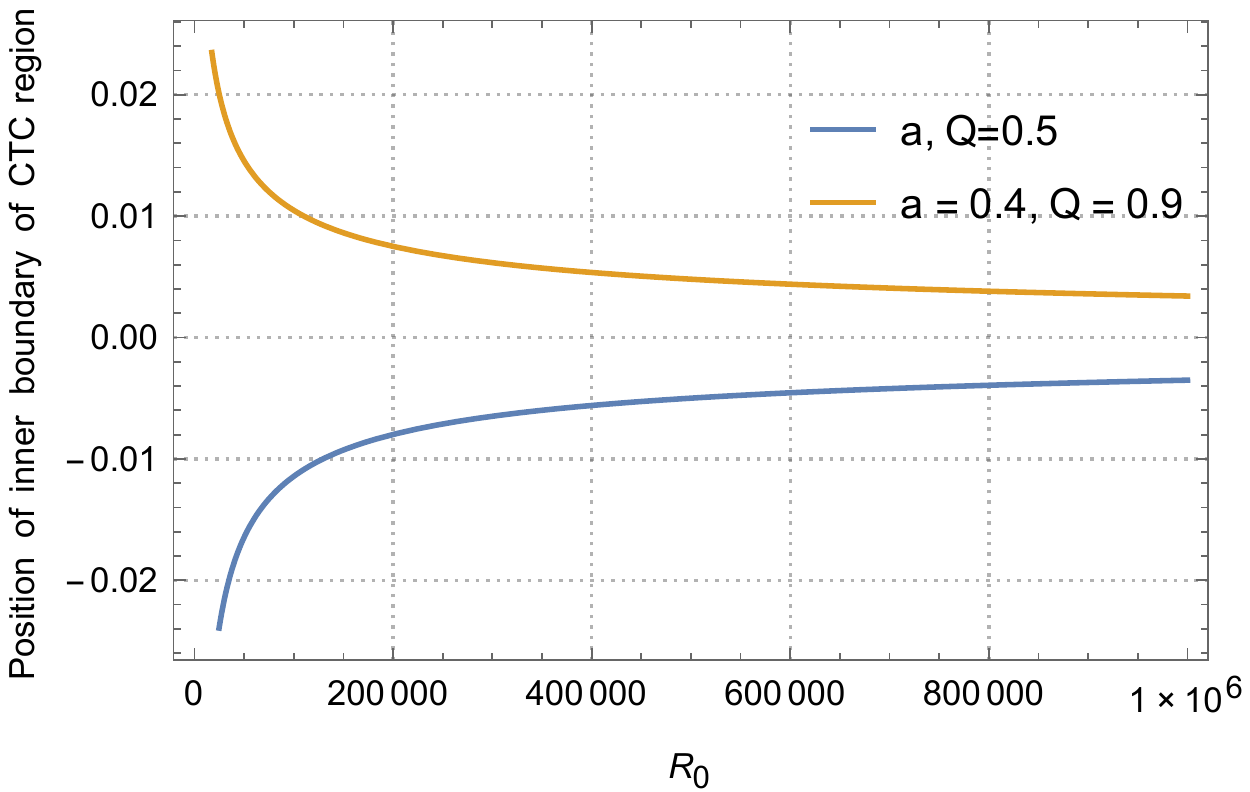}
\caption{Plot of the inner and outer roots of $g_{\phi\phi}$ vs. $x$. The top curve is for the inner root and the bottom for the outer root. It can be seen that the roots approach each other as $R_0$ increases.}
\end{figure*}
\section{\label{sec:disc}Results}

The key results of this paper are as follows. We revisited the problem of Closed Time-like Curves in Kerr-Newman BH and two other modified Kerr-Newman space-time/black-hole models, namely the Non-commutativity inspired KN BH and the KN BH from the f(R) gravity theory.

In the Non-commutativity inspired model of the Kerr-Newman Black-hole is a semi-classical model that effectively incorporates non-commutativity of space-time, resolving the singularity, replacing it with a Gaussian distribution. The model is parametized by the usual BH parameters ${M,a,Q}$ and the non-commutativity parameter $p$, the existence of the causality violating region ( the region containing closed time-like curves) region depends on the value of the non-commutativity parameter $p$. We graphically showed the existence of the CTC region for particular values of $p$. Thus, although the non-commutative model of the Kerr-Newman black hole has effectively resolved the problems connected with the singularity, it does not help in preserving causality. We then analyzed geodesic turning points for null geodesics in the CTC region and have shown that turning points can exist in the CTC region, and thus, causality violating information can be carried away from the CTC region to the outside asymptotically flat space-time region as they are causally connected. However, we also found that the non-commutativity parameter $p$ can be chosen such that the region containing closed time-like curves vanishes. The critical value of $p=p_{cr}$ for a Black hole for which the CTC region degenerates to its boundary depends on its BH parameters: the Mass, specific charge and specific angular momentum ${M,a,q}$. The region within the inner horizon thus makes a transition from a viscious set to a regular region without a Cauchy horizon at the critical value of the non commutative parameter. The exact nature of this transition is left for future considerations.

We also analyzed the causality violating region of the f(R) Kerr-Newman Black-hole. The main feature of the $f(R)$ theory being that the modified Einstein's equation admits a non-zero Ricci scalar $R_0$ as a vacuum solution. Thus, here we also studied the dependence of the size of the CTC region on $R_0$. We discussed by explicit examples, the existence of the CTC region and its size. We then analyzed geodesic turning points in the CTC region for null geodesics and found that, even this case, the region is causally connected with the asymptotically flat region. Also, similar to the situation in NC models, the Ricci scalar $R_0$ can be chosen such that no causality violating region exists. The citical value of $R_0$ for which this happens is shown to be only dependent on the mass of the BH in contrast to the previous case. Since the $f(R)$ models of Black holes with a constant $R_0$ are equivalent to de Sitter/Anti de Sitter Kerr-Newman black holes of Einstein's theory, the results derived here are also applicable to them. In the last section, we discussed the Kerr-Newman dS black hole corresponding to the present day inferred value of the cosmological constant, showing the existence of the causality violating region. We then show that, for a super massive black hole of mass $\sim 10^{12} M_{\odot}$, a ridiculously high value of the (effective) cosmological constant  $R_0$ would be necessary to make the CTC region disappear .

The causality violating region and it's charactarization and quantifying parameter range for it's elimination is important from various prespectives. The causality violaton in KN-deSitter spacetime is directly relevant from the point of view of our cosmological data that indicates a positive cosmological constant. The relevance of the interior region of KN-deSitter spacetime can be argued  from  \citep{cardoso18}, \citep{hintz2016}, \citep{hintz2016dec}. The extent of causality violation and it's impact can be fully understood through a better understanding of the process of gravitational collapse that forms a Kerr Newman black hole. For proving or disproving the Chronology Protection Conjecture, one needs to understand, through numerical or theoretical means, the sequence of major events during the process of gravitational collapse. This can help us decide which assumptions and features of modified spacetime solutions are desirable from the point of view of causality protection. We may be able to understand some of the features concerned with gravitational collapse, singularity fromation and the dynamics of the interior region of black holes from the emitted gravitational waves from the system. These can hint us in further progress in undestanding Chornology protecting mechanisms.  
With the increased sophistication expected in gravitational wave astronomy one might eventually discern the features/information of causality violating regions, if they exist, from the formation of one or the merger of blackholes.

\section*{Acknowledgements}
We would like to acknowledge IUCAA and BITS, Pilani - Hyderabad Campus for the resources and computational facilities.  
\bibliography{mybib.bib}

\begin{thebibliography}{10}

\bibitem{abbott2016}
B.~P. et~al Abbott.
\newblock Observation of gravitational waves from a binary black hole merger.
\newblock {\em Phys. Rev. Lett.}, 116:061102, Feb 2016.

\bibitem{hawkingandellis}
S.~W. Hawking and George~F. Ellis.
\newblock {\em The large scale structure of space-time [by] S. W. Hawking and
  G. F. R. Ellis}.
\newblock University Press Cambridge [Eng.], 1973.

\bibitem{novikov1997}
I.~D. Novikov and V.~P. Frolov.
\newblock {\em Black hole Physics, Basic Concepts and New Developments}.
\newblock Springer, 1998.

\bibitem{bojowald2005}
Martin Bojowald, Rituparno Goswami, Roy Maartens, and Parampreet Singh.
\newblock Black hole mass threshold from nonsingular quantum gravitational
  collapse.
\newblock {\em Phys. Rev. Lett.}, 95:091302, Aug 2005.

\bibitem{Mbonye2005}
Manasse~R. Mbonye and Demosthenes Kazanas.
\newblock Nonsingular black hole model as a possible end product of
  gravitational collapse.
\newblock {\em Phys. Rev. D}, 72:024016, Jul 2005.

\bibitem{hossenfelder2010}
Sabine Hossenfelder, Leonardo Modesto, and Isabeau Pr\'emont-Schwarz.
\newblock Model for nonsingular black hole collapse and evaporation.
\newblock {\em Phys. Rev. D}, 81:044036, Feb 2010.

\bibitem{Note1}
In some cases also extends 'beyond' the singularity to negative radii for e.g.
  Kerr ST, which may be considered physically irrelevant.

\bibitem{cardoso18}
Vitor Cardoso, Jo\~ao~L. Costa, Kyriakos Destounis, Peter Hintz, and Aron
  Jansen.
\newblock Quasinormal modes and strong cosmic censorship.
\newblock {\em Phys. Rev. Lett.}, 120:031103, Jan 2018.

\bibitem{pankaj2011}
Pankaj~S. Joshi and Daniele Malafarina.
\newblock Recent developments in gravitational collapse and spacetime
  singularities.
\newblock {\em International Journal of Modern Physics D}, 20(14):2641--2729,
  2011.

\bibitem{malafarina2017}
Daniele Malafarina.
\newblock Classical collapse to black holes and quantum bounces: A review.
\newblock {\em Universe}, 3(4):48, May 2017.

\bibitem{nathanail17}
Antonios Nathanail, Elias~R. Most, and Luciano Rezzolla.
\newblock Gravitational collapse to a kerr–newman black hole.
\newblock {\em Monthly Notices of the Royal Astronomical Society: Letters},
  469(1):L31--L35, 2017.

\bibitem{hawking91}
S.~W. Hawking.
\newblock Chronology protection conjecture.
\newblock {\em Phys. Rev. D}, 46:603--611, Jul 1992.

\bibitem{godel1949}
Kurt G\"odel.
\newblock An example of a new type of cosmological solutions of einstein's
  field equations of gravitation.
\newblock {\em Rev. Mod. Phys.}, 21:447--450, Jul 1949.

\bibitem{morris1988}
Michael~S. Morris, Kip~S. Thorne, and Ulvi Yurtsever.
\newblock Wormholes, time machines, and the weak energy condition.
\newblock {\em Phys. Rev. Lett.}, 61:1446--1449, Sep 1988.

\bibitem{carter1968}
Brandon Carter.
\newblock Global structure of the kerr family of gravitational fields.
\newblock {\em Phys. Rev.}, 174:1559--1571, Oct 1968.

\bibitem{mthesis}
Vaishak Prasad.
\newblock The causal structure of kerr-newman spacetime.
\newblock Master's thesis, BITS,Pilani Hyderabad Campus, 2015.

\bibitem{defelice1980}
F~de~Felice, L~Nobili, and M~Calvani.
\newblock Charged singularities: The causality violation.
\newblock {\em Journal of Physics A: Mathematical and General}, 13(12):3635,
  1980.

\bibitem{Slobodov2008}
Sergei Slobodov.
\newblock Unwrapping closed timelike curves.
\newblock {\em Foundations of Physics}, 38(12):1082, Oct 2008.

\bibitem{gonzalez1996}
Pedro~F. Gonz\'alez-D\'{\i}az.
\newblock Ringholes and closed timelike curves.
\newblock {\em Phys. Rev. D}, 54:6122--6131, Nov 1996.

\bibitem{salazar17}
J.~Felix Salazar and Thomas Zannias.
\newblock Behavior of causal geodesics on a kerr\char21{}de sitter spacetime.
\newblock {\em Phys. Rev. D}, 96:024061, Jul 2017.

\bibitem{boulware1992}
David~G. Boulware.
\newblock Quantum field theory in spaces with closed timelike curves.
\newblock {\em Phys. Rev. D}, 46:4421--4441, Nov 1992.

\bibitem{lmodesto10}
Leonardo Modesto and Piero Nicolini.
\newblock Charged rotating noncommutative black holes.
\newblock {\em Phys. Rev. D}, 82:104035, Nov 2010.

\bibitem{cembranos14}
J.~A.~R. Cembranos, A.~de~la Cruz-Dombriz, and P.~Jimeno Romero.
\newblock Kerr–newman black holes in f(r) theories.
\newblock {\em International Journal of Geometric Methods in Modern Physics},
  11(01):1450001, 2014.

\bibitem{nicolini2006}
Piero Nicolini, Anais Smailagic, and Euro Spallucci.
\newblock Noncommutative geometry inspired schwarzschild black hole.
\newblock {\em Physics Letters B}, 632(4):547 -- 551, 2006.

\bibitem{ansoldi2007}
Stefano Ansoldi, Piero Nicolini, Anais Smailagic, and Euro Spallucci.
\newblock Non-commutative geometry inspired charged black holes.
\newblock {\em Physics Letters B}, 645(2):261 -- 266, 2007.

\bibitem{spallucci2009}
Euro Spallucci, Anais Smailagic, and Piero Nicolini.
\newblock Non-commutative geometry inspired higher-dimensional charged black
  holes.
\newblock {\em Physics Letters B}, 670(4):449 -- 454, 2009.

\bibitem{nicolini2005}
Piero Nicolini.
\newblock A model of radiating black hole in noncommutative geometry.
\newblock {\em Journal of Physics A: Mathematical and General}, 38(39):L631,
  2005.

\bibitem{hackmann2013}
Eva Hackmann and Hongxiao Xu.
\newblock Charged particle motion in kerr-newmann space-times.
\newblock {\em Phys. Rev. D}, 87:124030, Jun 2013.

\bibitem{saheb16}
Saheb Soroushfar, Reza Saffari, Sobhan Kazempour, Saskia Grunau, and Jutta
  Kunz.
\newblock Detailed study of geodesics in the kerr-newman-(a)ds spacetime and
  the rotating charged black hole spacetime in $f(r)$ gravity.
\newblock {\em Phys. Rev. D}, 94:024052, Jul 2016.

\bibitem{Note2}
There are other actions that lead to the same EOM.

\bibitem{Note3}
The boundary term is to be added to complete the gravitational action.

\bibitem{hintz2016}
P.~Hintz and A.~Vasy.

\bibitem{hintz2016dec}
P.~Hintz.

\end{thebibliography}
\end{document}